\documentclass[%
 reprint,
 amsmath,amssymb,
 aps,
]{revtex4-2}
\usepackage{float}
\usepackage{physics,amsmath}
\usepackage{graphicx}
\usepackage{dcolumn}
\usepackage{bm}
\usepackage{hyperref}
\usepackage[mathlines]{lineno}

\begin{document}

\preprint{npjQI/Preprint}

\title{Fixed-Time Gaussian State Transfer via Collective Dissipation in a Fully Static Architecture}

\author{Austen Couvertier}
\affiliation{%
Physics Department, Stevens Institute of Technology, Hoboken, New Jersey 07030, USA
}%
 \email{acouvert@stevens.edu}
\author{Ting Yu}%
 \email{tyu1@stevens.edu}
\affiliation{%
Physics Department, Stevens Institute of Technology, Hoboken, New Jersey 07030, USA
}%

\date{\today}

\begin{abstract}
\textbf{Abstract.}
We establish a fully static, dissipation-only primitive for Gaussian state transfer in which collective coupling to a shared environment fixes a single transfer time without coherent transport or time-dependent control. The dynamics close in a bright–dark decomposition, yielding an analytic transfer time \(t^{*}=\pi/g_{a}\) and a real interference factor that eliminates dynamical phase accumulation, so that all phase sensitivity arises solely from the Gaussian fidelity metric. This defines a fixed-time Gaussian communication channel in which dissipation mediates information transfer without active control. Ensemble analysis shows that channel performance remains robust to amplitude-level asymmetries, which act as uniform-rate dressing, whereas phase-level system detuning disrupts the fixed-time mechanism via dynamical phase winding and produces revival behavior. Incorporating finite Ornstein–Uhlenbeck memory via a zeroth-order O-operator expansion yields only rate and phase renormalization without introducing new dynamical pathways. In the fast-memory regime, environmental correlations are rapidly suppressed, rendering the channel insensitive to environmental detuning and restoring the Markovian limit. These results show that collective dissipation alone is sufficient to support fixed-time Gaussian state transfer, challenging the assumption that coherent transport or active control are required for continuous-variable communication.
\end{abstract}

\maketitle


\section{Introduction}\label{sec:intro}

Quantum state transfer is a central primitive in quantum communication and distributed information processing \cite{cirac_quantum_1997,pellizzari_quantum_1997,bose_quantum_2003}. Continuous-variable Gaussian systems provide an analytically tractable setting with a well-developed information-theoretic framework \cite{braunstein_quantum_2005,weedbrook_gaussian_2012,eisert_gaussian_2005}. Rather than serving as intermediary buses \cite{pirandola_quantum_2006}, bosonic modes here act as both carriers and medium, emphasizing operational constraints over platform generality. \cite{adesso_entanglement_2007,van_loock_multipartite_2000}.

Existing state-transfer schemes achieve high fidelity by relaxing operational constraints through control, transport, or engineered dissipation. These include time-dependent control \cite{parkins_quantum_1999,stannigel_optomechanical_2011}, transport through extended channels \cite{vermersch_quantum_2017,rai_transfer_2022}, interference-based suppression of dissipation \cite{lau_high-fidelity_2019,luo_optimal_2025}, and engineered reservoirs \cite{kraus_preparation_2008,diehl_quantum_2008}. Environment-assisted protocols similarly rely on incorporating additional resources into the design \cite{plenio_dephasing-assisted_2008,pirandola_environment-assisted_2021,wang_passive_2025}. These approaches leave open the question of whether a fully static, dissipation-only architecture can support a timed transfer primitive. Here, we realize such a mechanism as a fixed-time Gaussian channel mapping sender to receiver under purely dissipative dynamics.

All couplings are static, with no external drives, measurement, feedback, or engineered reservoirs. Collective dissipation arises solely from a shared environment \cite{barreiro_open-system_2011}. In this setting, the Lindblad channel itself mediates transfer at a fixed time determined by dissipative dynamics. No coherent transport, adiabatic protocol, or engineered dissipation is employed \cite{kraus_preparation_2008,diehl_quantum_2008,verstraete_quantum_2008}.

This setting differs from decoherence-free subspace (DFS) approaches, which preserve states \cite{an_entanglement_2005} by encoding them into noise-invariant subspaces \cite{lidar_decoherence-free_2003}. Here, dissipation does not protect coherence but actively generates transfer at a fixed time. Unlike DFS encoding, this implements an active dissipative transfer, placing it within the class of Gaussian communication channels.

Environmental structure modifies operating conditions without introducing control. In the Markovian limit, the bath contributes only background decay. Finite Ornstein–Uhlenbeck memory renormalizes bright-sector rates and introduces phase accumulation without generating new pathways \cite{plenio_dephasing-assisted_2008,rebentrost_environment-assisted_2009,cheng_optomechanical_2019, huelga_non-markovianity-assisted_2012, li_non-markovian_2020}. In the fast-memory regime, environmental correlations are rapidly suppressed, restoring effective Markovian channel behavior.  Thus, memory deforms the effective channel parameters while preserving the primitive’s structure \cite{rivas_quantum_2014,tamascelli_nonperturbative_2018,mui_enhanced_2025,wang_robust_2023}. From an information-theoretic perspective, this corresponds to controlled deformation of channel parameters without altering channel structure.

We consider a minimal model with bilinear coupling and a single collective Lindblad channel. The task is fixed-time recovery of an arbitrary Gaussian input state, with performance quantified by Gaussian fidelity \cite{banchi_quantum_2015}. The dynamics close in a bright–dark basis, yielding analytic expressions for the moments and the interference structure that enables transfer. Finite memory enters via an O-operator \cite{yu_non-markovian_1999} treatment that modifies effective rates without introducing new pathways.

We establish a minimal setting in which collective dissipation alone supports fixed-time Gaussian state transfer. The mechanism arises from bright–dark interference, yielding analytic predictions for transfer time and dynamics. We characterize fidelity as a function of Gaussian parameters and quantify robustness to system asymmetries. Finite memory produces rate and phase renormalization without altering the transfer mechanism. These results define the operating regime of a dissipation-only transfer channel and provide the basis for the analysis that follows.

\section{Method}\label{sec:method}

\subsection{System definition and operating regime}
\label{subsec:system_definition}

We consider a reduced bosonic system comprising two photon modes, $(a,c)$, and two phonon modes, $(b,d)$, coupled by static, time-independent interactions. The system consists of two identical subsystems, each defined by interacting photon-phonon pairs: $a$ and $b$ in the first, and $c$ and $d$ in the second. The interaction is modeled by a generic bilinear photon-phonon interaction, which for one subsystem (e.g., $a$ and $b$) is given by the full bilinear form,
\begin{equation}\label{eq:H0_generic}
    \hat{H}_0/\hbar = \omega_{a}\hat{a}^{\dagger}\hat{a}+\omega_{b}\hat{b}^{\dagger}\hat{b} + g_{a}(\hat{b}+\hat{b}^{\dagger})(\hat{a}e^{i\phi_a} + \hat{a}^{\dagger}e^{-i\phi_a}),
\end{equation}
which contains both excitation-conserving and counter-rotating contributions. This form captures common linearized photon-phonon interactions in cavity QED and optomechanics \cite{parkins_quantum_1999,aspelmeyer_cavity_2014}.

We study the operating regime where a rotating-wave approximation (RWA) applies, i.e., $g_a, |\delta_{ab}|\ll\omega_{a,b}$. Therefore, the system reduces to an excitation-conserving effective Hamiltonian for each local photon-phonon pair,
\begin{equation*}
    \hat{H}_{a,b}/\hbar = \delta_{ab}\hat{b}^{\dagger}\hat{b} + g_{a}(\hat{a}\hat{b}^{\dagger}e^{i\phi_a} + \mathrm{h.c.}).
\end{equation*}
An analogous expression, $\hat{H}_{c,d}$ defines the second subsystem $(c,d)$ where $a,b\to c,d$. We assume identical modes, $\omega_a=\omega_c$ and $\omega_b=\omega_d$, implying $\delta_{ab}=\delta_{cd}$. The total system Hamiltonian is given by the sum of the two local contributions,
\begin{equation*}
    \hat{H}_{\mathrm{sys}} = \hat{H}_{a,b} + \hat{H}_{c,d},
\end{equation*}
and allows us to state the task of this work:
\begin{center}
    \noindent\fbox{
    \begin{minipage}{0.95\linewidth}
    \textbf{Task}: Implement fixed-time transfer of a quantum state in the sender mode ($b$) at time $t^*$, to the target mode ($d$). In this work, we restrict to dynamics that preserve Gaussianity, so the sender mode is always initialized in a single-mode Gaussian state, while all other modes are initialized as vacuum states.    
    \end{minipage}
    }
\end{center}
The choice of $t^*$ is determined by the open-system mechanism introduced in Sec.~\ref{subsec:collective_dissipation}. We define a symmetric reference configuration with $\delta_{ab}=0$, $g_c=g_a$, and $\phi_c=\phi_a$. Deviations are introduced via detuning $\delta_{ab}$, coupling imbalance $g_c=g_a(1+\eta)$, and phase mismatch $\delta\phi=\phi_c-\phi_a$, treated as static, time-independent, imperfections.

\subsection{Collective mode structure of the closed system}
\label{subsec:collective_structure}

To expose the pathways made available by symmetry (i.e., in the symmetric reference configuration  $\delta_{ab}=\eta=\delta\phi=0$), it is convenient to express the photon degrees of freedom in a collective basis defined by symmetric and antisymmetric linear combinations of the bare photon modes:
\begin{equation}\label{eq:photon_bright_dark_definition}
    \hat{m}_{\pm} = \tfrac{1}{\sqrt{2}}(\hat{a} \pm \hat{c}).
\end{equation}
Analogously, the phonon degrees of freedom can be recast,
\begin{equation}\label{eq:phonon_bright_dark_definition}
\hat{n}_\pm = \frac{1}{\sqrt{2}}( \hat{b} \pm \hat{d} ).
\end{equation}
In the symmetric reference configuration defined in Sec.~\ref{subsec:system_definition}, the system Hamiltonian acquires a block structure when written in these collective bases. The resulting Hamiltonian reduces to a sum of two independent collective sectors:
\begin{equation}\label{eq:Hsys_bright_dark}
     \hat{H}_{\mathrm{sys}}'/\hbar =  g_a(\hat{m}_{-}\hat{n}_{-}^{\dagger}+ \hat{m}_{+}\hat{n}_{+}^{\dagger}) +\mathrm{h.c.},
\end{equation}
with $\mathrm{h.c.}$ being the hermitian conjugate of the shown terms. At the level of the closed-system Hamiltonian, these sectors evolve as:
\begin{align}\label{eq:closed_ODEs}
    \tfrac{d}{dt}\langle \hat{m}_{\pm}\rangle &= -i\,g_a\langle \hat{n}_{\pm}\rangle\\
    \tfrac{d}{dt}\langle \hat{n}_{\pm}\rangle &= -i\,g_a\langle \hat{m}_{\pm}\rangle.
\end{align}
This collective decomposition identifies two distinct structural pathways by which information initially encoded in a single phonon mode redistributes among collective degrees of freedom. However, at the Hamiltonian level alone, this structure does not imply state transfer between subsystems. Symmetry enforces identical evolution of $\hat{n}_\pm$, leaving the target mode $\hat{d}$ unpopulated. (Appendix \ref{app:deterministic_transfer_time}).

\subsection{Open-system description: Markovian collective dissipation}
\label{subsec:collective_dissipation}

The transfer mechanism only emerges once the system is coupled to an environment that distinguishes between the collective photon modes. What follows models the environment as a continuum of bosonic modes linearly coupled to the photon degrees of freedom,
\begin{equation}\label{eq:H_env_definition}
    \hat{H}_{\mathrm{env}}/\hbar = \sum_{\bm{\lambda}}\omega_{\bm{\lambda}}\hat{a}_{\bm{\lambda}}^{\dagger}\hat{a}_{\bm{\lambda}} + g_{\bm{\lambda}}(\hat{L}\hat{a}_{\bm{\lambda}}^{\dagger} + \hat{L}^{\dagger}\hat{a}_{\bm{\lambda}}).
\end{equation}
At the level of the reduced dynamics, photon dissipation is captured by a single collective jump operator,
\begin{equation}\label{eq:L_collective_dissipation}
    \hat L = \sqrt{\kappa}\,(\hat a + \sqrt{1+\epsilon}\,\hat c).
\end{equation}
where a static asymmetry parameter ($\epsilon$) allows for unequal coupling of the two photon modes to the environment. We neglect phonon decoherence and local photon loss to isolate collective dissipation. In the symmetric limit, the jump operator becomes proportional to the symmetric collective photon mode,
\begin{equation*}
    \hat L = \sqrt{\kappa}\,(\hat a +\hat c) = \sqrt{2\kappa}\hat{m}_{+},
\end{equation*}
defining $\hat{m}_+$ as a bright mode, while $\hat{m}_-$ remains dark. Assuming weak system-environment coupling and bath correlation times short compared to all system timescales, the reduced moment dynamics are governed by a Markovian adjoint Lindblad equation generated by the system Hamiltonian and the collective jump operator\cite{breuer_theory_2009},
\begin{equation}\label{eq:Markov_ME}
    \frac{d}{dt}\langle \hat{A} \rangle = -\frac{i}{\hbar}\langle[\hat{A},\hat{H}_{sys}]\rangle + \frac{1}{2}\langle \hat{L}^{\dagger}[\hat{A},\hat{L}] + [\hat{L}^{\dagger},\hat{A}]\hat{L} \rangle.
\end{equation}
The dynamics define the first-order ODEs below, 
\begin{align}\label{eq:first_order_ODEs}
    \tfrac{d}{dt}\langle \hat{m}_{-}\rangle &= -i\,g_a\langle \hat{n}_{-}\rangle \qquad \tfrac{d}{dt}\langle \hat{m}_{+}\rangle = -i\,g_a\langle \hat{n}_{+}\rangle-\kappa\langle \hat{m}_{+}\rangle\\
    \tfrac{d}{dt}\langle \hat{n}_{-}\rangle &= -i\,g_a\langle \hat{m}_{-}\rangle \qquad \tfrac{d}{dt}\langle \hat{n}_{+}\rangle = -i\,g_a\langle \hat{m}_{+}\rangle.
\end{align}
This collection highlights the open-system asymmetry that defines the transfer primitive study in this work:
\begin{center}
    \noindent\fbox{
    \begin{minipage}{0.95\linewidth}
    \textbf{Primitive}:  
    A combination of the symmetric system ($\hat{H}_{\mathrm{sys}}$) and collective dissipation ($\hat{L} \propto\hat{m}_{+}$) that, for $\pi g_a/\kappa \ll 1$ at $t^*=\pi/g_a$, enables state transfer from sender ($b$) to the target ($d$), defining a fixed-time dissipative channel between the two modes.
    \end{minipage}
    }
\end{center}
In the weak-coupling regime $g_a/\kappa\ll1$, the bright sector is overdamped with effective rate $\kappa_{\mathrm{eff}}=g_a^2/\kappa$, while the dark sector remains coherent. At $t^*=\pi/g_a$, the dark-mode inversion enables transfer provided $\pi g_a/\kappa\ll1$. A detailed derivation is provided in Appendix \ref{app:Markov_Exact}.

This defines the baseline state-transfer primitive studied in this work, which we refer to as the ``baseline primitive". Finite environmental memory modifies the effective rates governing these dynamics without altering the structure of the dissipation channel, as discussed in Sec.~\ref{subsec:open_system_models}.

\subsection{Finite-memory environment model}
\label{subsec:open_system_models}

Finite environmental memory is treated as a perturbation to the baseline primitive. This study models finite memory using an Ornstein-Uhlenbeck (OU) bath characterized by a single-pole colored-noise correlation function,
\begin{equation}\label{eq:OU_correlation_function}
    \alpha(t,s) =  \frac{\gamma}{2}e^{-(\gamma+i\delta_{\mathrm{env}})(t-s)} \qquad \text{for } t\geq s
\end{equation}
with $\gamma^{-1}$ setting the bath memory timescale and $\delta_{\mathrm{env}} = \Omega-\omega_a$ the system-environment detuning for an environmental central frequency $\Omega$. 

The OU environment is treated using non-Markovian quantum state diffusion with a zeroth-order O-operator (O-zero) truncation \cite{diosi_non-markovian_1998,strunz_open_1999,yu_non-markovian_1999}. At the $O_0$ level, memory enters through a time-dependent operator, $\hat{\bar{O}}_{0}(t)$, in a time-local master equation. The details and derivation of this operator are provided in Appendix \ref{app:oOperator} such that system dynamics evolve under an updated time-local master equation: 
\begin{equation}\label{eq:OU_ME}
    \frac{d}{dt}\langle \hat{A} \rangle = -\frac{i}{\hbar}\langle[\hat{A},\hat{H}_{sys}]\rangle + \langle \hat{\bar{O}}_{0}^{\dagger}(t)[\hat{A},\hat{L}] + [\hat{L}^{\dagger},\hat{A}]\hat{\bar{O}}_{0}(t) \rangle.
\end{equation}
OU dynamics are obtained by integrating the time-local $O_0$ equations with kernel $\alpha(t,s)$. In the weak-coupling regime, memory effects manifest as effective renormalizations of the bright-sector dynamics without introducing new pathways, as provided in Appendix \ref{app:ouRates}.

\subsection{Gaussian states and phase-space description}
\label{subsec:gaussian_phase_space}

The open system dynamics defined in Secs.~\ref{subsec:system_definition}-\ref{subsec:open_system_models}, namely a bilinear Hamiltonian and linear Lindblad operator, define a Gaussian preserving map \cite{weedbrook_gaussian_2012,ferraro_gaussian_2005}. We, therefore, restrict our study to single-mode Gaussian inputs and track the first and second moments.

The sender phonon mode $b$ is initialized in a displaced--squeezed vacuum,
\begin{equation}\label{eq:initial_state_definition}
    \ket{\psi}_0 = (\hat{\mathcal{D}}(\beta)\hat{\mathcal{S}}(r_\beta)\ket{0}_{b})\otimes\ket{0}_{d,m_{\pm}},
\end{equation}
with displacement $\beta \in \mathbb{C}$ and squeezing $r_{\beta} \in \mathbb{C}$. The phases $\Theta=\arg(\beta)$ and $\Phi=\arg(r_{\beta})$ characterize the input state in phase space. All other modes $(d,m_{\pm})$ are initialized in vacuum.

We adopt standard phase-space quadratures for each mode $o\in\{b,d,m_{\pm}\}$,
\begin{equation}\label{eq:quadrature_definition}
    \hat{q}_o = \frac{1}{\sqrt{2}}(\hat{o}+\hat{o}^{\dagger}) \qquad \hat{p}_o = \frac{1}{i\sqrt{2}}(\hat{o}-\hat{o}^{\dagger}),
\end{equation}
with associated quadrature vectors $\hat R_o=(\hat q_o,\hat p_o)^T$ and covariance matrices,
\begin{equation}\label{eq:covariance_definition}
    (\sigma_o)_{i,j} = \tfrac{1}{2}\langle\{(\Delta \hat{R}_o)_i,(\Delta \hat{R}_o)_j\}\rangle \qquad i\in\{1,2\},
\end{equation}
with centered moments $\Delta \hat{R}_o = \hat{R}_o - \langle \hat{R}_o\rangle$, the anti-commutator $\{,\}$, and $i$ as the quadrature vector index. \cite{weedbrook_gaussian_2012,ferraro_gaussian_2005}. The trajectories of these moments are used to evaluate the fidelity metric \cite{banchi_quantum_2015}.

\subsection{Gaussian fidelity and transfer metric}
\label{subsec:gaussian_fidelity}

Transfer performance is evaluated using the single-mode Gaussian fidelity following Banchi \emph{et al.}~\cite{banchi_quantum_2015},
\begin{multline}\label{eq:Banchi_Fidelity}
    \mathcal{F}_t =\mathcal{F}(\{\langle\hat{R}_b\rangle_0,\sigma_b^0\},\{\langle\hat{R}_d\rangle_t,\sigma_d^t\}) \\= \mathrm{F}(\sigma_b^0,\sigma_d^t)e^{-\tfrac{1}{4}(\delta^{T}_t(\sigma_b^0+\sigma_d^t)^{-1}\delta_t)}.
\end{multline}
Where,
\begin{equation}\label{eq:Banchi_Prefactor}
    \mathrm{F}(\sigma_b^0,\sigma_d^t) = (\sqrt{\Lambda_t+\bm{\Delta}_t}-\sqrt{\Lambda_t})^{-1/4},
\end{equation}
\begin{align}\label{eq:Banchi_Invariants}
     \Lambda_t = 4\,\mathrm{det}[&\sigma_b^0+\tfrac{i}{2}\bm{\Omega}]\mathrm{det}[\sigma_d^t+\tfrac{i}{2}\bm{\Omega}]\\ \bm{\Delta}_t &= \mathrm{det}[\sigma_b^0+\sigma_d^t]\nonumber,
\end{align}
\begin{equation}\label{eq:Gaussian_Symplectic}
    \bm{\Omega} =
    \begin{pmatrix}0&1\\-1&0\end{pmatrix},
\end{equation}
as defined in Ref.~\cite{banchi_quantum_2015}. We report squared fidelity $\mathcal{F}_t^2$ to enable comparison with other state-transfer primitives. Additionally, we can define the success condition as: 
\begin{center}
    \noindent\fbox{
    \begin{minipage}{0.95\linewidth}
    \textbf{Objective}:  
    Transfer the sender state to the target at $t^* = \pi/g_a$, defining $\mathcal{F}_{t^*}^2 > 0.95$ as high-fidelity channel transmission.
    \end{minipage}
    }
\end{center}
We define an admissible set, $\mathcal{S}$, of accessible states that exclude inputs with $\mathcal{F}_{t=0}^2 \geq 0.95$. The admissible set's definition is dependent on the allowable parameter ranges for the sender mode \cite{aspelmeyer_cavity_2014}, which we defined as $\mathcal{B}$ such that
\begin{align}\label{eq:parameter_space}
    \mathcal{B} = \{(|\beta|,|r_\beta|,2\Theta-\Phi) \mid 0&\leq|\beta|\leq1 \,\wedge\,0\leq|r_\beta|\leq1.5 \nonumber\\
    &\wedge\,-\pi\leq2\Theta-\Phi\leq\pi\}
\end{align}
where $2\Theta-\Phi$ is a displacement-squeeze misalignment that is an explicit dependence of the fidelity measure \cite{banchi_quantum_2015}. Finally, we define the admissible set using the initial fidelity restriction:
\begin{equation}\label{eq:admissible_set}
    \mathcal{S} = \{(|\beta|,|r_\beta|,2\Theta-\Phi) \in \mathcal{B} \mid \mathcal{F}_{t=0}^2 < 0.95\}.
\end{equation}
States are sampled uniformly over $\mathcal{S}$. This sample set is used when reporting statistics on the robustness of the primitive for states that initially lie outside the success threshold.

\subsection{Numerical procedures and parameter regimes}
\label{subsec:numerics}

The first- and second-moment equations for the collection $\{\hat{m}_{\pm},\hat{b},\hat{d}\}$ are integrated directly in the Heisenberg picture for both Markovian and OU environments using the adjoint evolution equations introduced in Secs.~\ref{subsec:collective_dissipation} and ~\ref{subsec:open_system_models}, respectively.

Markovian dynamics give time-independent moment dynamics, while OU dynamics introduce explicit time-dependence through the dressed operator $\hat{\bar O}_{0}(t)$ as in Sec.~\ref{subsec:open_system_models}. Furthermore, the creation operator coefficients of $\hat{\bar O}_{0}(t)$ are exactly 0 (Appendix \ref{app:oOperator_Obar}). The ODE set is reduced accordingly, and the resulting moment trajectories are processed using the same fidelity evaluation procedure as in the Markovian case.

In all simulations, physicality is enforced using the covariance-based uncertainty relation, $\sigma_o^{t} - \frac{i}{2}\bm{\Omega} \geq 0$ for $o\in\{b,d\}$, which is numerically satisfied throughout integration. Gaussian squared-fidelity is then evaluated from the numerically obtained first and second moments. The readout time, $t^{*}=\pi/g_a$, is fixed by the fully symmetric reference configuration as defined in Appendix \ref{app:Markov_Exact} and is not varied to maximize any of the perturbative results. Additionally, the capture window ($\Delta t$), is defined as 0 whenever $\mathcal{F}_{t^*}^2 < 0.95$ and as $\Delta t = t_{+} - t_{\mathrm{-}}$ whenever $\mathcal{F}_{t^*}^2 > 0.95$. $t_{\pm}$ define the time interval around $t^*$ such that $\mathcal{F}_{t}^2 \geq 0.95$ for all $t \in [t_-,t_+]$. Additionally, we limit the time interval such that $0<t_{-}<t^*<t_{+}\leq2\,t^{*}$. This is motivated by the study's restriction: all metrics are fixed to the symmetric configuration and not optimized. 

The system and environmental parameters explored in this work are provided in the following table. Parameter variations are implemented as systematic sweeps of the open-system equations and do not correspond to time-dependent controls.
\begin{center}
\begin{tabular}{ |c|c|c| } 
\hline
Parameter & Range & Default \\
\hline
$\delta_{ab}/\kappa$ & $[-0.05,0.05]$ & 0\\
$\delta_{\mathrm{env}}/\kappa$ & $[-2,2]$ & 0.6\\
$\gamma/\kappa$ & $[0.5,5.0]$ & 0.5\\
$g_{a}/\kappa$ & $[0.005,0.05]$ & 0.02\\
$\eta$ & $[-0.2,0.2]$ & 0\\
$\epsilon$ & $[-0.2,0.2]$ & 0\\
$|r_{\beta}|$ & $[0.0,1.5]$ & 1\\
$|\beta|$ & $[0,1.0]$ & 1\\
$(2\Theta-\Phi)/\pi$ & $[-1,1]$ & 0\\
\hline
\end{tabular}
\end{center}

\section{Results}\label{sec:results}

\subsection{Primitive Baseline: Markovian Environment Under Symmetric Conditions}
\label{subsec:Results_Baseline}

In the symmetric Markovian limit, the closed-form moment dynamics reduce to a single real interference factor $f(t)$ governing all target-mode quadratures as detailed in Appendix~\ref{app:Markov_Exact}. In the weak-coupling regime $g_{a}/\kappa\ll 1$, this yields: (i) a \emph{fixed, state-independent} transfer time $t^{*}=\pi/g_{a}$, and (ii) the absence of dynamical phase accumulation. The analytic fidelity at $t^{*}$ therefore depends only on $f(t^{*})$, independent of input state. Operationally, this defines a Gaussian channel with a universal decoding time that is independent of the input-state parameters.
\begin{figure}
    \centering
    \includegraphics[width=\linewidth]{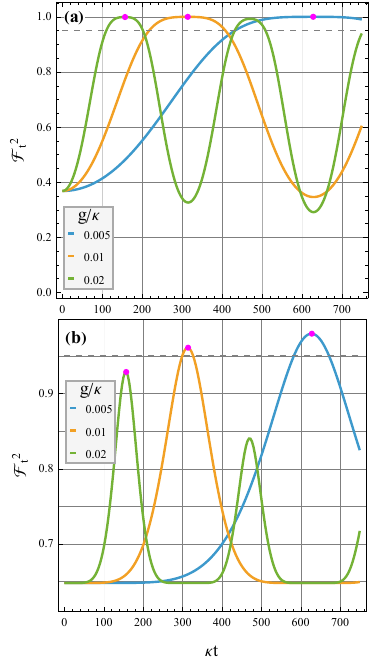}
    \caption{Baseline primitive under symmetric Markovian conditions. Time-resolved squared Gaussian fidelity $\mathcal{F}_t^{\,2}$ for three coupling ratios $g_a/\kappa = 0.005,\,0.01,\,0.02$. Panel (a) uses a displaced-vacuum sender state $(|\beta|=1, r_\beta=0)$; panel (b) uses a squeezed-vacuum sender state $(|\beta|=0, r_\beta=1)$. The curves show the full numerical Markovian dynamics as functions of the scaled time $\kappa t$. Magenta markers denote the squared Gaussian fidelity at analytic transfer time $t^{*}=\pi/g_a$, evaluated for each $g_a/\kappa$. The dashed horizontal line indicates the success threshold $\mathcal{F}_{t}^{\,2}=0.95$. For small $g_a/\kappa$, the peak of $\mathcal{F}_t^{\,2}$ aligns closely with the analytic prediction at $t^{*}$; for larger $g_a/\kappa$, bright-sector damping reduces the achievable peak fidelity but the primitive’s fixed-time structure remains evident. The displaced and squeezed cases exhibit the same timing behavior, with differences arising only from the distinct first- and second-moment structures of the input Gaussian states.}
    \label{fig:PrimitiveConfirmation}
\end{figure}
Figure~\ref{fig:PrimitiveConfirmation} confirms the predicted fixed-time transfer across representative Gaussian inputs. The numerical fidelity maxima coincide with the analytic prediction (magenta dots) at $t^{*}=\pi/g_a$, demonstrating state-independent transfer timing. Deviations at larger $g_a/\kappa$ are governed by bright-sector damping, indicating amplitude loss as the dominant limitation.

\begin{figure}
    \centering
    \includegraphics[width=\linewidth]{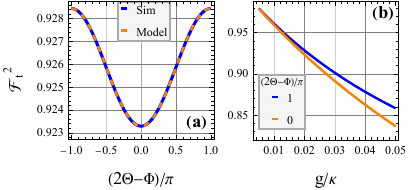}
    \caption{Phase dependence of the Gaussian fidelity metric at the fixed readout time. (a) Squared fidelity, $\mathcal{F}_{t^{*}}^{2}$, as a function of sender misalignment $(2\Theta-\Phi)/\pi$ for a representative coupling ratio $g)a/\kappa=0.02$. Numerical results (blue) are compared directly with the analytic expression derived from the closed-form fidelity structure (orange). (b) Dependence of $\mathcal{F}_{t^{*}}^{2}$ on the coupling ratio $g_a/\kappa$ for two representative input-state orientations: aligned ($2\Theta-\Phi=\pi$, blue) and misaligned ($2\Theta-\Phi=0$, orange). Both panels use fixed state magnitudes $(|\beta|,|r_\beta|)=(1,1)$. The observed cosine-shaped variation with respect to $(2\Theta-\Phi)$ and the consistent analytic–numerical agreement confirm that all phase sensitivity arises from the intrinsic Gaussian fidelity metric, while the transfer dynamics themselves introduce no additional phase accumulation.}
    \label{fig:MetricPhaseDependence}
\end{figure}

Because phase sensitivity is isolated in the fidelity metric, we test whether the dynamics introduce any additional phase rotation at $t^{*}$. This motivates Fig.~\ref{fig:MetricPhaseDependence}, which examines whether the observed fidelity variation with respect to the sender’s phase-space orientation arises solely from the metric or whether dynamical phase accumulation plays a role.
Taken with Appendix~\ref{app:analytic_metric}, this indicates that the channel does not introduce dynamical phase noise. Figure~\ref{fig:MetricPhaseDependence} shows that all phase dependence arises from the Gaussian fidelity metric, with numerical results matching the analytic prediction across the full parameter range. No additional phase rotation or distortion is introduced by the dynamics, confirming that the channel does not generate dynamical phase noise. The greater degradation of misaligned states reflects the amplification of metric sensitivity by bright-sector loss.

\begin{figure}
    \centering
    \includegraphics[width=\linewidth]{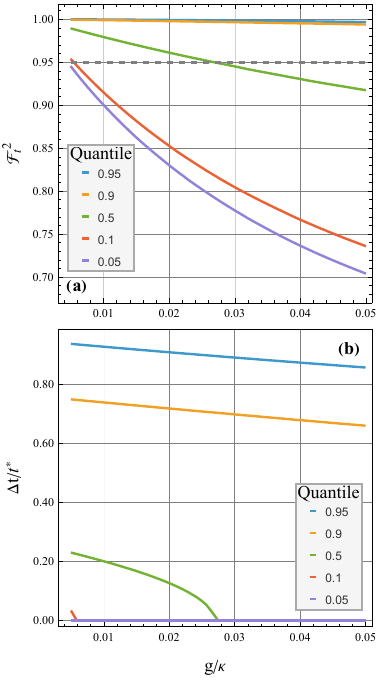}
    \caption{Input-ensemble channel performance of the baseline primitive. Quantile statistics of (a) the squared fidelity, $\mathcal{F}_{t^{*}}^{2}$, and (b) the normalized capture window, $\Delta t/t^{*}$, evaluated at the fixed readout time $t^{*}=\pi/g_a$ for 3374 randomly sampled sender mode initializations from the accessible set $\mathcal{S}$ (after rejection from 3500 randomly sampled sender modes from set $\mathcal{B}$). Each curve reports quantiles computed independently for the fidelity and capture-window distributions. As $g_a/\kappa$ increases, fidelity quantiles exhibit a gradual downward shift, while capture-window quantiles show a corresponding narrowing, reflecting the increased influence of bright-sector decay. The quantile spread remains modest for the 0.5 - 0.95 quantiles as $g_a/\kappa$ increases, indicating that the primitive’s fixed-time transfer mechanism is robust over half the Gaussian input states in $\mathcal{S}$. The lower percentiles (0.1 and 0.05) are more sensitive to increasing $g_a/\kappa$ in a similar fashion as the misalignment in Fig.~\ref{fig:MetricPhaseDependence}.}
    \label{fig:PrimitiveEnsemble}
\end{figure}

These results together establish fixed-time transfer without dynamical phase accumulation. With these properties verified, we next characterize the primitive’s performance over the full admissible input-ensemble $\mathcal{S}$ defined in Sec.~\ref{subsec:gaussian_fidelity}. This is reported in Fig.~\ref{fig:PrimitiveEnsemble}, which quantifies both fidelity and timing robustness across thousands of Gaussian inputs.
Figure~\ref{fig:PrimitiveEnsemble} shows smooth degradation of channel performance with increasing $g/\kappa$, while high-fidelity transfer is maintained across most of the admissible ensemble in the weak-coupling regime. The spread across quantiles reflects input-state dependence as measured by the fidelity metric, while the underlying transfer mechanism remains robust. The capture window narrows with increasing $g_a/\kappa$, reflecting reduced temporal tolerance as bright-sector loss increases. 

Overall, the primitive remains robust across a broad class of Gaussian inputs, with performance limited by bright-sector loss and input-state misalignment. We next examine the effect of static asymmetries.

\subsection{Primitive Robustness: System \& Markovian Asymmetries}
\label{subsec:Results_Markov_Asymmetries}

We examine amplitude-level asymmetries in coupling ($\eta$) and dissipation ($\epsilon$), both of which induce bright–dark mixing and introduce decoherence into the transfer channel. 

\begin{figure*}
    \centering
    \includegraphics[width=\linewidth]{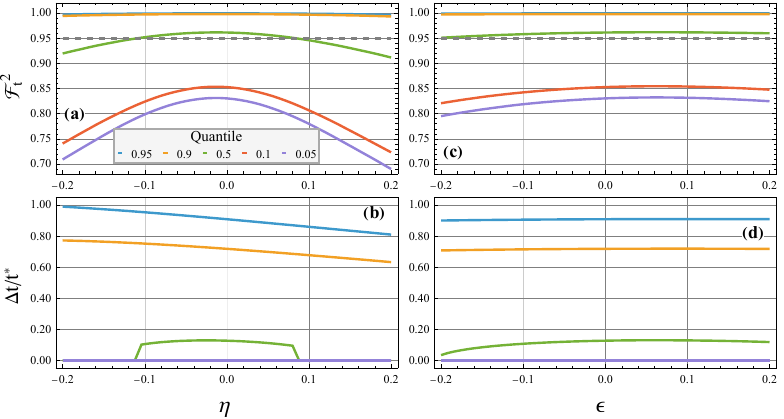}
    \caption{Sensitivity of the primitive to amplitude-level system asymmetries. (a,b) Ensemble quantiles of the squared fidelity $\mathcal{F}_{t^{*}}^{2}$ and normalized capture window $\Delta t/t^{*}$ under a coupling-magnitude imbalance $\eta$. (c,d) The corresponding quantities under a decay-rate imbalance $\epsilon$. Curves are computed at the fixed readout time $t^{*}=\pi/g_{a}$ with coupling ratio $g_a/\kappa=0.02$ using the same ensemble $\mathcal{S}$ as in Fig.~\ref{fig:PrimitiveEnsemble}. Both imperfections introduce bright-dark mixing, destroying the underlying DFS. This directly decreases fidelity and temporal tolerance, but does so smoothly and without inducing phase-accumulation effects or dynamical instabilities. The individual degradations are asymmetric between the two parameters: transfer survives and even benefits for slightly negative $\eta$ and slightly positive $\epsilon$. These correspond to weaker overall coupling and stronger overall decay, respectively. This further reduces the coupling-decay ratio in both cases, explaining the shifts as stronger adherence to the weak-coupling regime before asymmetries introduce mode mixing and system-wide decoherence.}
    \label{fig:AmplitudeAsymmetries}
\end{figure*}

\begin{figure*}
    \centering
    \includegraphics[width=\textwidth]{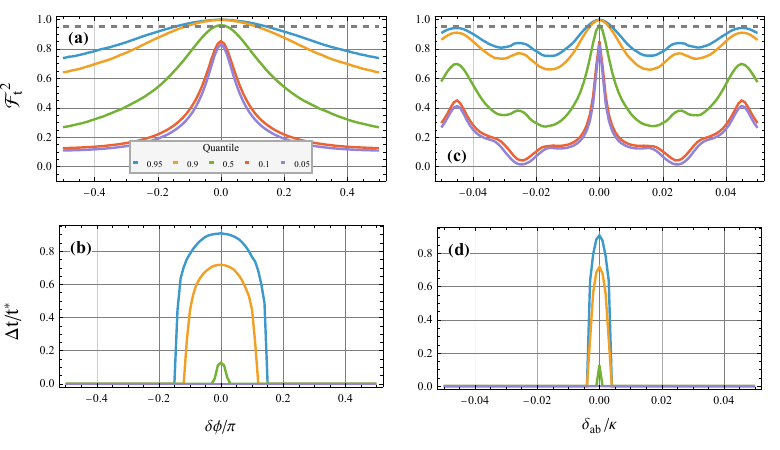}
    \caption{Effects of phase-level system asymmetries on the primitive. (a,b) Quantile curves of $\mathcal{F}_{t^{*}}^{2}$ and $\Delta t/t^{*}$ versus the coupling-phase offset $\delta\phi/\pi$. Because $\delta\phi$ only rotates the phonon collective basis [Appendix~\ref{app:coupling_phase}], the degradation is monotonic, symmetric, and free of revival structure. (c,d) The same quantities versus phonon–photon detuning $\delta_{ab}/\kappa$ on a narrow interval around resonance. Unlike $\delta\phi$, detuning modifies both dark- and bright-sector rates, generating rapid dynamical phase accumulation and producing the oscillatory “revival’’ structure in fidelity and capture window whenever the accumulated phase realigns at $t^{*}$.}
    \label{fig:PhaseAsymmetries}
\end{figure*}

\begin{figure}
    \centering
    \includegraphics[width=\linewidth]{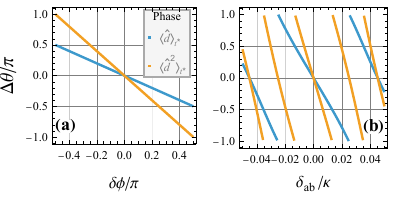}
    \caption{Phase accumulation induced by phase-level asymmetries. (a) Phase of the first-order moment $\mathrm{arg}[\langle\hat d\rangle_{t^{*}}]$ and second-order moment $\mathrm{arg}[\langle\hat d^{2}\rangle_{t^{*}}]$ evaluated at the fixed readout time under a coupling-phase offset $\delta\phi/\pi$. Both grow (or decrease) linearly with $\delta\phi$, consistent with a global unitary rotation. (b) Corresponding phases under phonon–photon detuning $\delta_{ab}/\kappa$, showing rapid nonlinear variation and repeated $2\pi$ windings even for small detunings, in alignment with the revival structure observed in Fig.~\ref{fig:PhaseAsymmetries}(c,d).}
    \label{fig:MarkovPhaseAccumulation}
\end{figure}

Figure~\ref{fig:AmplitudeAsymmetries} shows that both asymmetries produce a smooth, monotonic degradation of channel performance. The effect is dominated by increased decoherence, which suppresses the interference factor governing transfer. Upper quantiles remain near unity for moderate asymmetries, while lower quantiles degrade more rapidly due to increased metric sensitivity. The capture window narrows with increasing asymmetry, reflecting reduced temporal tolerance under enhanced decoherence. This indicates that amplitude asymmetries act as uniform channel noise rather than inducing dynamical instabilities.
Importantly, neither asymmetry introduces dynamical phase accumulation, and their effects are confined to amplitude-level channel degradation. The situation changes markedly when considering phase-level asymmetries, for which the dynamics can develop a rapid phase mismatch between collective sectors. This is explored in Fig.~\ref{fig:PhaseAsymmetries}, followed by explicit phase-accumulation diagnostics in Fig.~\ref{fig:MarkovPhaseAccumulation}.
Phase-level asymmetries introduce dynamical phase mismatch between collective modes, modifying the interference structure at the fixed readout time, $t^{*}=\pi/g_{a}$. Two distinct mechanisms are relevant: a global coupling-phase offset $\delta\phi = \phi_c - \phi_a $, and a phonon–photon detuning $\delta_{ab} = \omega_{b(d)}-\omega_{a(c)} $, with analytic treatments summarized in Appendix~\ref{app:Markov_Asymmetries}.

Figure~\ref{fig:PhaseAsymmetries}(a–b) shows that a global coupling-phase offset produces smooth, monotonic degradation of fidelity without revival structure, as it induces only a uniform phase rotation. The resulting performance loss arises entirely from metric misalignment rather than dynamical modification of the transfer mechanism.

In contrast, phonon–photon detuning modifies both dark- and bright-sector dynamics. Fig.~\ref{fig:PhaseAsymmetries}(c–d) shows time-dependent phase accumulation and shifting effective transfer rates. This produces oscillatory fidelity behavior with multiple revival peaks as a function of detuning. These revivals arise from dynamical phase wrapping: fidelity is restored when the accumulated phase at $t^*$ realigns modulo $2\pi$ as shown by the intersections in Fig.~\ref{fig:MarkovPhaseAccumulation}. This corresponds to periodic rephasing of the effective channel.
These realignment points are consistent with the phase-accumulation profiles in Fig.~\ref{fig:MarkovPhaseAccumulation}(b), which show that even small detunings generate rapid, nonlinear phase variation on the scale of $10^{-2}\kappa$. 

Amplitude asymmetries act as uniform channel noise, reducing fidelity without altering the transfer mechanism. Phase asymmetries, by contrast, introduce dynamical phase mismatch, with detuning producing strong oscillatory sensitivity through phase accumulation. These effects identify detuning as the dominant instability of the Markovian primitive. We now extend the analysis to include finite environmental memory.

\subsection{Primitive Robustness: Environmental Correlations}
\label{subsec:Results_OU_Asymmetries}

We now examine the effect of finite environmental memory using an Ornstein–Uhlenbeck (OU) bath. In this setting, memory modifies the bright-sector dynamics through time-dependent rate dressing without introducing new transfer pathways. A central question is whether a fast environment suppresses these memory-induced deviations and restores effective Markovian channel behavior.
\begin{figure*}
    \centering
    \includegraphics[width=\linewidth]{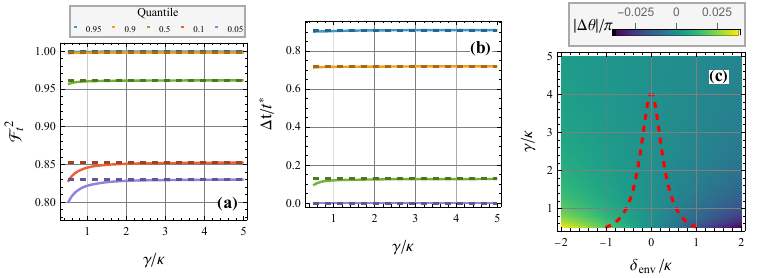}
    \caption{Memory–rate dependence of the OU environment. (a) Quantiles of the squared fidelity $\mathcal{F}_{t^{*}}^{2}$ as a function of the memory rate $\gamma/\kappa$, compared with the Markov reference (dashed). All quantiles begin below the Markov value when $\gamma/\kappa \lesssim 1$, and increase monotonically toward the Markov limit as $\gamma/\kappa$ grows, indicating rapid suppression of memory-induced deviations in the O-zero regime. (b) Corresponding quantiles of the normalized capture window $\Delta t/t^{*}$, showing the same monotonic convergence to the Markov baseline. (c) Phase accumulation $|\Delta\theta|/\pi$ at the fixed readout time over the $(\delta_{\mathrm{env}}/\kappa,\gamma/\kappa)$ plane for a displaced-vacuum probe state. Significant phase accumulation is confined to small $\gamma/\kappa$ and finite detuning, while increasing $\gamma/\kappa$ strongly suppresses the effect. The red dashed curve marks the approximate boundary of the "fast-memory regime" determined by the condition $|\mathrm{Im}(\gamma_{\mathrm{O}})| \approx 15\,g_{a}$.}
    \label{fig:OUDressing}
\end{figure*}
Figure~\ref{fig:OUDressing}(a,b) shows that finite memory produces a uniform degradation of channel performance at small $\gamma/\kappa$, followed by rapid monotonic convergence to the Markovian limit as $\gamma/\kappa$ increases. This behavior is consistent across all quantiles, indicating that memory acts as a global dressing of the effective channel parameters rather than introducing state-dependent effects, as confirmed in Appendix~\ref{app:ouRates}. This identifies a fast-memory regime in which environmental correlations are effectively quenched, restoring Markovian channel behavior. In this regime, the environment acts as a dynamical filter that suppresses memory-induced distortions at the channel level.

Fig.~\ref{fig:OUDressing}(c) shows that phase accumulation is confined to slow-memory, detuned regimes. As $\gamma/\kappa$ increases, the imaginary component of the dynamics is rapidly suppressed, and phase accumulation vanishes. The red-dashed boundary marks the onset of the fast-memory regime, where the channel becomes insensitive to environmental detuning. In this regime, memory-induced phase effects are dynamically filtered out. 

We next examine how environmental detuning modifies the channel under finite memory.
\begin{figure}
    \centering
    \includegraphics[width=\linewidth]{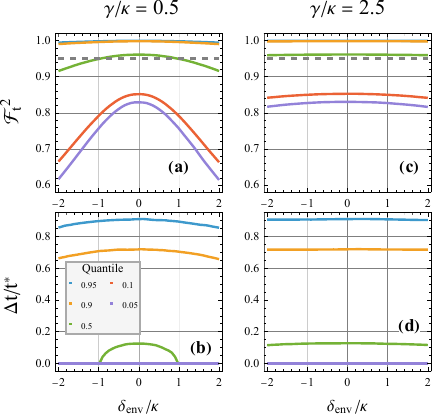}
    \caption{Effects of environmental detuning under finite OU memory. (a,b) Quantiles of the squared fidelity $\mathcal{F}_{t^{*}}^{2}$ and normalized capture window $\Delta t/t^{*}$ as functions of the environmental detuning $\delta_{\mathrm{env}}/\kappa$ for a long memory time scale $\gamma/\kappa=0.5$. Both quantities exhibit smooth, symmetric variation about resonance, with modest degradation at large detuning and no oscillatory or revival structure. (c,d) The same quantiles for a short-memory timescale, $\gamma/\kappa=2.5$, where all curves are nearly flat and coincide with their Markov baselines, indicating that environmental detuning becomes negligible for short memory times. Data are evaluated at the fixed readout time $t^{*}=\pi/g_{a}$ using the ensemble $\mathcal{S}$ defined in Fig.~\ref{fig:PrimitiveEnsemble}.}
    \label{fig:OUAsymmetry}
\end{figure}
Figure~\ref{fig:OUAsymmetry}(a,b) shows that under long memory times ($\gamma/\kappa=0.5$), environmental detuning produces smooth, symmetric degradation of channel performance without oscillatory structure. Fidelity is maximal near resonance and decreases monotonically with detuning, with stronger sensitivity for lower quantiles. Unlike in system detuning, no revival behavior is observed, indicating that environmental detuning does not induce coherent phase wrapping.

In the fast-memory regime ($\gamma/\kappa=2.5$) shown in Fig.~\ref{fig:OUAsymmetry}(c,d), all quantiles collapse onto their Markovian baselines, and the dependence on environmental detuning becomes negligible. This demonstrates that a fast environment suppresses both rate and phase distortions, effectively restoring the baseline channel up to small perturbative corrections.

Environmental detuning produces smooth rate dressing without inducing oscillatory or unstable behavior, in contrast to system detuning. Its influence is rapidly suppressed as the memory rate increases, highlighting the robustness of the primitive to bath-spectral imperfections. Only system-level asymmetries directly disrupt the coherent transfer mechanism.

\section{Discussion}\label{sec:discussion}

We establish a fully static, dissipation-only mechanism for Gaussian state transfer distinct from transport-based architectures. Unlike schemes relying on propagation, routing, or pulse shaping \cite{zeng_quantum_1994,wang_using_2012,dong_optomechanical_2012,tian_adiabatic_2012,rodriguez-lara_propagation_2014,rezaei_accelerated_2022,molinares_transfer_2023,navarathna_continuous_2023,hahto_transfer_2025,rakhubovsky_photon-phonon-photon_2017,luo_optimal_2025}, the present mechanism requires no coherent transport or control. Transfer arises from symmetry-protected bright–dark interference, recovering an arbitrary Gaussian state at a fixed time set by static system parameters. Because no excitations propagate, performance is governed by internal timing rather than channel length or topology. This defines a dissipative Gaussian communication channel governed entirely by internal dynamics.

The primitive operates without measurement, feedback, or active control \cite{lloyd_quantum_1999,andersen_unconditional_2005,pfister_continuous-variable_2020,ortiz-gutierrez_continuous_2017,anuradha_perfect_2025,wang_autonomous_2019}. With static couplings and a single collective dissipation channel, the dynamics close in a bright–dark basis and are governed by a real interference factor \(f(t)\). This yields a state-independent transfer time \(t^{*}=\pi/g_{a}\) with dark-mode inversion and overdamped bright dynamics. Because \(f(t^{*})\) is real, the dynamics introduce no phase noise, and all phase sensitivity arises from the Gaussian metric. Ensemble analysis shows uniform channel behavior across Gaussian inputs, with variation determined by metric alignment rather than dynamics.

Amplitude-level asymmetries ($\eta$,$\epsilon$) introduce bright–dark mixing that renormalizes effective rates without generating new frequencies or phase dynamics. They produce smooth, monotonic degradation of fidelity and temporal tolerance, with minor shifts when the coupling–decay ratio is effectively reduced. The absence of oscillations indicates that amplitude asymmetries act as uniform channel noise without destabilizing the transfer mechanism.

Phase asymmetries introduce dynamical phase mismatch between collective modes. A coupling-phase offset ($\delta\phi$) induces only a global phase rotation, causing smooth fidelity degradation without affecting the transfer mechanism. In contrast, phonon–photon detuning ($\delta_{ab}$) modifies dynamical rates and introduces phase accumulation, producing oscillatory revivals through phase wrapping. This identifies frequency matching as the dominant stability constraint.

Environmental memory enters as a renormalization of bright-sector rates without introducing new dynamical pathways \cite{wang_passive_2025, lacroix_non-markovian_2024}. As a result, memory cannot generate a phase winding or revival structure and preserves the underlying bright–dark interference mechanism \cite{li_non-markovian_2020,laine_nonlocal_2012}. In the long-memory regime (small \(\gamma/\kappa\)), performance degrades smoothly, and phase shifts appear only when both memory and detuning are present \cite{huelga_non-markovianity-assisted_2012}. In the fast-memory regime, environmental correlations are rapidly quenched, restoring Markovian channel behavior. As \(\gamma/\kappa\) increases, both rate and phase distortions are suppressed. The channel becomes insensitive to environmental detuning, effectively filtering memory-induced phase and rate distortions. 

The perturbative $O_0$ treatment excludes strong memory, thermal noise, non-Gaussian inputs, and network extensions. The fixed transfer time is intrinsic and cannot be optimized without introducing control. The primitive is single-shot and therefore lacks error correction. The required conditions align with cavity QED, optomechanical, and microwave–mechanical platforms, in which collective dissipation arises naturally \cite{palomaki_state_2013,aspelmeyer_cavity_2014,weaver_coherent_2017,lake_two-colour_2020,zhong_microwave_2022}. Finite memory is intrinsic to realistic baths and does not require engineered reservoirs. In this setting, the mechanism defines a baseline static communication element that can be combined with active control or error-mitigation strategies when required. These results show that dissipation can serve as a structural component of a communication channel rather than solely as a source of noise. The framework extends directly to broader dissipation-mediated Gaussian information tasks.

\appendix

\section{Baseline Primitive Derivation}
\label{app:Markov_Exact}

This appendix provides the technical derivation underlying the baseline primitive results reported in Secs.~\ref{subsec:Results_Baseline}. We repeat this restricts the derivation to the fully symmetric configuration $\delta_{ab}=\eta=\epsilon=\delta\phi=0$ and a white-noise Markovian environment. The system dynamics are governed by the system Hamiltonian, bath coupling, and adjoint master equation defined in Eqs.~\eqref{eq:Hsys_bright_dark}–\eqref{eq:Markov_ME}. Initial Gaussian state and fidelity metric are defined in Eq.~\eqref{eq:initial_state_definition} and Eqs.~\eqref{eq:Banchi_Fidelity}-\eqref{eq:Gaussian_Symplectic}.

\subsection{First-order moment from the bright-dark basis}
\label{app:markov_first_order_exact}

Substituting the bright–dark operators into the adjoint master equation~\eqref{eq:Markov_ME} yields a closed linear system for the first-order moments $\langle\hat m_{\pm}\rangle$ and $\langle\hat n_{\pm}\rangle$. The compact form given by Eq.~\eqref{eq:first_order_ODEs} in the Sec.~\ref{subsec:collective_dissipation}. The initial condition~\eqref{eq:initial_state_definition} fixes $\langle\hat m_{\pm}\rangle_0=0$ and $\langle\hat n_{\pm}\rangle_0=\beta/\sqrt{2}$. Solving the dark-sector block of Eq.~\eqref{eq:first_order_ODEs} yields undamped oscillatory solutions
\begin{equation}\label{eq:first_order_dark_phonon}
    \langle\hat{n}_{-}\rangle_t = \frac{\beta}{\sqrt{2}}\,\cos(g_a\,t) = \frac{\beta}{\sqrt{2}}\,\mathrm{T}_{-}(t).
\end{equation}
Solving the bright-sector block of Eq.~\eqref{eq:first_order_ODEs} yields a \emph{damped} differential equation,
\begin{equation}\label{eq:first_order_bright_phonon_ODE}
    \tfrac{d^2}{dt^2}\langle\hat{n}_{+}\rangle = -\kappa\tfrac{d}{dt}\langle\hat{n}_{+}\rangle-g_a^2\langle \hat{n}_{+}\rangle,
\end{equation}
with an initial derivative $\tfrac{d\langle\hat{n}_{+}\rangle}{dt}|_0= 0$. The resulting dynamics of the bright phonon mode are given as
\begin{equation}\label{eq:first_order_bright_phonon}
     \langle\hat{n}_{+}\rangle_t = \frac{\beta}{\sqrt{2}}\,e^{-\kappa\,t/2}[\cosh(\tfrac{\tilde{g}}{2}t)+\frac{\kappa}{\tilde{g}}\sinh(\tfrac{\tilde{g}}{2}t)] = \frac{\beta}{\sqrt{2}}\,\mathrm{T}_{+}(t),
\end{equation}
with an effective frequency $\tilde{g} = \sqrt{\kappa^2 - 4g_a^2}$. In the overdamped regime, $2g_a <  \kappa$, the bright mode evolves as stated because $\tilde{g} \in \mathbb{R}$. Once $2g_a > \kappa$, the functions are analytically continued by the oscillatory counterparts. This breaks the primitive's underlying mechanism and prevents the separation of timescales discussed in Sec.~\ref{subsec:collective_dissipation}. 

These results combine to define the target mode's first-order moment evolution as,
\begin{align}\label{eq:exact_target_first_order_moment}
    \langle \hat{d} \rangle_t &=\frac{1}{\sqrt{2}}(\langle \hat{n}_{+}\rangle_t - \langle \hat{n}_{-}\rangle_t)\nonumber\\ &= \frac{\beta}{2}[\mathrm{T_{+}}(t)-\mathrm{T_{-}}(t)]\nonumber\\
    &= \frac{\beta}{2}f(t) = \frac{\langle \hat{b} \rangle_0}{2}f(t),
\end{align}
which takes the form of a time-dependent scalar function, $f(t)$. For real-valued system parameters ($g_a,\kappa\in\mathbb{R}$), $f(t)\in\mathbb{R}$ is a real-valued time-dependent function for all times.

\subsection{Second-order moments from bright-dark basis}
\label{app:markov_second_order_exact}

The remaining second-order moments can likewise be derived from the combinations of the independent channels, where
\begin{align}
    \langle\hat{d}^{2}\rangle &= \frac{1}{2}(\langle \hat{n}_{+}^2\rangle - 2\langle \hat{n}_{+}\hat{n}_{-}\rangle + \langle \hat{n}_{-}^2\rangle) \\
    \langle\hat{d}^{\dagger}\hat{d}\rangle &= \frac{1}{2}(\langle \hat{n}_{+}^\dagger\hat{n}_{+}\rangle - 2\mathbb{R}[\langle \hat{n}_{+}\hat{n}_{-}^{\dagger}\rangle] + \langle \hat{n}_{-}^\dagger\hat{n}_{-}\rangle)
\end{align}
define exactly how the bright-dark structure relates to the physical target mode. 

Beginning with $\langle \hat{d}^2 \rangle$, the time evolution defined by Eq.~\eqref{eq:Markov_ME} divides the required operators into three independent ODE sets grouped as:
\begin{align*}
    \langle \hat{n}_{-}^2 \rangle&\to\{\langle \hat{n}_{-}^2 \rangle,\langle \hat{m}_{-}\hat{n}_{-} \rangle,\langle \hat{m}_{-}^2 \rangle,\mathrm{h.c.}\}, \\
    \langle \hat{n}_{+}\hat{n}_{-}\rangle&\to\{\langle \hat{n}_{+}\hat{n}_{-}\rangle,\langle \hat{m}_{-}\hat{n}_{+}\rangle,\langle \hat{m}_{+}\hat{n}_{-}\rangle,\langle \hat{m}_{-}\hat{m}_{+}\rangle,\mathrm{h.c.}\}, \\
    \langle \hat{n}_{+}^2 \rangle&\to\{\langle \hat{n}_{+}^2\rangle,\langle \hat{m}_{+}\hat{n}_{+}\rangle,\langle \hat{m}_{+}^2\rangle,\mathrm{h.c.}\}, \\
\end{align*}
with $\mathrm{h.c.}$ defined as the hermitian conjugate of all operators mentioned in the set. Given the initial conditions, assuming all modes except the $b$ mode are initialized to vacuum, normally ordered second-order moments involving the $b$ mode are non-zero while all other normally ordered second-order moments are zero. From the above mentioned sets, only $\langle\hat{n}_{-}^2\rangle_0 = \langle\hat{n}_{+}\hat{n}_{-}\rangle_0 = \langle\hat{n}_{+}^2\rangle_0 =\langle\hat{b}^2\rangle_0/2$, with all other operators initialized at 0. The resulting ODEs defined by Eq.~\eqref{eq:Markov_ME} are linear and take the same form as the first-order moments. This admits closed-form solutions that are also proportional to the appropriate product of $\mathrm{T}_{\pm}$. In the case of the moments defining $\langle\hat{d}^2\rangle$, we report
\begin{align}
    \langle \hat{n}_{-}^2 \rangle_t &= \frac{\langle \hat{b}^2 \rangle_0}{2}\mathrm{T}_{-}^{2}(t) \\
    \langle \hat{n}_{+}\hat{n}_{-} \rangle_t &= \frac{\langle \hat{b}^2 \rangle_0}{2}\mathrm{T}_{+}(t)\mathrm{T}_{-}(t) \\
    \langle \hat{n}_{+}^2 \rangle_t &= \frac{\langle \hat{b}^2 \rangle_0}{2}\mathrm{T}_{+}^{2}(t).
\end{align}
These combine to define the time-dependence of the second-order moment, $\langle \hat{d}^2\rangle_t$, using the previous definition
\begin{align}
    \langle \hat{d}^{2} \rangle_t &= \frac{1}{2}(\langle \hat{n}_{+}^2\rangle - 2\langle \hat{n}_{+}\hat{n}_{-}\rangle + \langle \hat{n}_{-}^2\rangle)\nonumber\\ &= \frac{\langle \hat{b}^2 \rangle_0}{4}(\mathrm{T}_{+}(t)-\mathrm{T}_{-}(t))^2\nonumber\\ &= \frac{\langle \hat{b}^2 \rangle_0}{4}f^2(t).
\end{align}
A similar framework defines the evolution of the moments that define $\langle\hat{d}^{\dagger}\hat{d}\rangle$, where the dynamics are divided into three independent ODE groups, namely
\begin{align*}
    \langle \hat{n}_{-}^\dagger\hat{n}_{-} \rangle&\to\{\langle  \hat{n}_{-}^\dagger\hat{n}_{-} \rangle,\langle  \hat{m}_{-}\hat{n}_{-}^\dagger \rangle,\langle \hat{m}_{-}^\dagger\hat{m}_{-} \rangle,\mathrm{h.c.}\}, \\
    \langle \hat{n}_{+}\hat{n}_{-}^{\dagger}\rangle&\to\{\langle \hat{n}_{+}\hat{n}_{-}^{\dagger}\rangle,\langle \hat{m}_{+}\hat{n}_{-}^{\dagger}\rangle,\langle \hat{m}_{-}\hat{m}_{+}^{\dagger}\rangle,\langle \hat{m}_{-}\hat{n}_{+}^{\dagger}\rangle,\mathrm{h.c.}\}, \\
    \langle \hat{n}_{+}^\dagger\hat{n}_{+} \rangle&\to\{\langle  \hat{n}_{+}^\dagger\hat{n}_{+} \rangle,\langle  \hat{m}_{+}\hat{n}_{+}^\dagger \rangle,\langle \hat{m}_{+}^\dagger\hat{m}_{+} \rangle,\mathrm{h.c.}\}.
\end{align*}
The initial conditions for these groups are $\langle \hat{n}_{-}^\dagger\hat{n}_{-} \rangle_0 = \langle \hat{n}_{+}^{\dagger}\hat{n}_{-}\rangle_0 = \langle \hat{n}_{+}^\dagger\hat{n}_{+} \rangle_0 = \langle \hat{b}^\dagger\hat{b} \rangle_0/2$. The solution set takes a form identical to the previous results, 
\begin{align}
    \langle \hat{n}_{-}^{\dagger}\hat{n}_{-} \rangle_t &= \frac{\langle \hat{b}^\dagger\hat{b} \rangle_0}{2}\mathrm{T}_{-}^{2}(t) \\
    \mathbb{R}[\langle \hat{n}_{+}\hat{n}_{-}^\dagger \rangle_t] &= \frac{\langle \hat{b}^\dagger\hat{b} \rangle_0}{2}\mathrm{T}_{+}(t)\mathrm{T}_{-}(t) \\
    \langle \hat{n}_{+}^{\dagger}\hat{n}_{+} \rangle_t &= \frac{\langle \hat{b}^\dagger\hat{b} \rangle_0}{2}\mathrm{T}_{+}^{2}(t),
\end{align}
with the only change being the sender moment that acts as the prefactor. These results, paired with the definition of $\langle\hat{d}^{\dagger}\hat{d}\rangle$, result in a form identical to $\langle\hat{d}^{2}\rangle$,:
\begin{equation}
    \langle \hat{d}^\dagger\hat{d} \rangle_t=\frac{\langle \hat{b}^\dagger\hat{b} \rangle_0}{4}f^2(t).
\end{equation}
In all cases, the target-mode moments evolve as a fixed proportion of the sender-mode moments, multiplied by a real-valued, time-dependent function determined solely by system-environment parameters. Dissipation is present in $f(t)$ via $\mathrm{T}_{+}(t)$, which encodes the inevitable decay of the bright phonon mode.

\subsection{State-transfer in the weak coupling limit}
\label{app:deterministic_transfer_time}

The target mode, being defined by its first and second moments, evolves deterministically via the derived equations
\begin{align*}
    \langle \hat{d} \rangle_t = \frac{\langle\hat{b} \rangle_0}{2}f(t&) \qquad 
    \langle \hat{d}^{2} \rangle_t = \frac{\langle\hat{b}^2 \rangle_0}{4}f^2(t) \\
    \langle \hat{d}^\dagger\hat{d} \rangle_t &= \frac{\langle \hat{b}^\dagger\hat{b} \rangle_0}{4}f^2(t)
\end{align*}
which depend on the initialized sender moments ($\langle\hat{b} \rangle_0, \langle\hat{b}^2 \rangle_0,\langle \hat{b}^\dagger\hat{b} \rangle_0$) and an identical time dependence
\begin{equation*}
    f(t) = e^{-\kappa\,t/2}[\cosh(\tfrac{\tilde{g}}{2}t)+\frac{\kappa}{\tilde{g}}\sinh(\tfrac{\tilde{g}}{2}t)] - \cos(g_a\,t)
\end{equation*}
defined by the photon-phonon coupling ($g_a$), collective dissipation ($\kappa$), and effective bright-mode coupling ($\tilde{g} = \sqrt{\kappa^2 - 4g_a^2}$). 

This overall time-dependence explains why dissipation is necessary. Taking the limit $\kappa \to 0$ for our task initialization, it follows that $\tilde{g} = 2i\,g_a$ and $\mathrm{T}_{+}(t) = \cos(g_a\,t)$ to give $f(t) = 0$ for all times. As $f(t)$ is the time-dependence for both first and second moments of the target-mode, we find $\langle \hat{d} \rangle_t=\langle \hat{d}^2 \rangle_t=\langle \hat{d}^\dagger\hat{d} \rangle_t=0$ for all $t$. This is the closed-system symmetry cancellation referenced in Sec.~\ref{subsec:collective_structure}.

With dissipation present, $\kappa \neq 0$, the analytic forms make it apparent that perfect mode matching requires $f(t) = 2$; in practice, dissipation prevents exact equality, and we instead seek a fixed readout time, $t^*$, where $f(t^*) \approx 2$, corresponding to a near-perfect state-transfer that defines the primitive. 

The form of $f(t)$ simplifies greatly in the weak-coupling ($g_a\ll\kappa$) regime, where the effective bright-mode coupling can be approximated as
\begin{equation}
    \tilde{g} = \kappa\sqrt{1-4(\frac{g_a}{\kappa})^2} \approx \kappa - \frac{2g_a^2}{\kappa}
\end{equation}
which is accurate up to $\mathcal{O}[(g_a/\kappa)^4]$. This combines with the exponential representation of the hyperbolic functions to give the approximate time dependence
\begin{equation}\label{eq:weak_coupling_f}
    f(t) \approx e^{-g_a^2\,t/\kappa}(1+\frac{g_a^2}{\kappa^2})- \frac{g_a^2}{\kappa^2}e^{-\kappa(1-\frac{g_a^2}{\kappa^2})t} -\cos(g_a\,t).
\end{equation}
In this form, the effective bright-mode contribution experiences an effective decay rate of $g_a^2/\kappa$. This allows for the bright dynamics to be effectively frozen on the timescale of the dark dynamics. We therefore use the dark dynamics to define the timing of the state-transfer, 
\begin{equation}\label{eq:analytic_transfer_time}
    t^{*} = \frac{\pi}{g_a}.
\end{equation}
When plugged into Eq.~\eqref{eq:weak_coupling_f}:
\begin{itemize}
    \item the dark contribution yields exactly $\cos(\pi) = -1$
    \item the first bright-term yields $e^{-\pi g_a/\kappa}(1+\frac{g_a^2}{\kappa^2}) \approx 1$ up to $\mathcal{O}[g_a/\kappa]$ in the operating regime $\pi g_a/\kappa \ll 1$ defined in Sec.~\ref{subsec:collective_dissipation}
    \item the second yields $\frac{g_a^2}{\kappa^2}e^{-\kappa\pi/g_a}e^{\pi g_a/\kappa} \approx 0$ when truncated to $\mathcal{O}[(g_a/\kappa)^2]$
\end{itemize}
Combining these three facts yields $f(t^*) \approx 2$ up to $\mathcal{O}[(g_a/\kappa)]$, fully defining the primitive studied in this work.

\subsection{Gaussian fidelity structure}
\label{app:analytic_metric}

The target-mode quadrature mean vector $\langle\hat R\rangle_{t}$ and covariance matrix $\sigma_{d}^{t}$ follow from Eqs.~\eqref{eq:quadrature_definition}–\eqref{eq:covariance_definition}. Substituting the analytic first- and second-order moments into the Gaussian root-fidelity expression~\eqref{eq:Banchi_Fidelity} yields the closed-form structure
\begin{equation}\label{eq:fidelity_struct}
    \mathcal{F}(|\beta|,|r_{\beta}|,\Theta,\Phi,t) = \mathrm{F}(|r_{\beta}|,t)e^{-\mathrm{A}(|\beta|,|r_{\beta}|,t)\,\mathrm{B}(|r_{\beta}|,\Theta,\Phi,t)}.
\end{equation}
Here $\mathcal{F}$ denotes the root-Uhlmann fidelity in the convention of Banchi \emph{et al.}; the squared fidelity $\mathcal{F}^2$ used for the success condition in Sec.~\ref{subsec:gaussian_fidelity} follows by squaring Eq. \eqref{eq:fidelity_struct}.

The terms $\mathrm{F}(|r_{\beta}|,t)$ and $\mathrm{A}(|\beta|,|r_{\beta}|,t)$ depend only on the time-dynamics, $f(t)$, and the sender-mode displacement and squeezing magnitudes ($|\beta|,|r_{\beta}|$). All phase dependence enters exclusively through $\mathrm{B}(|r_{\beta}|,\Theta,\Phi,t)$, which appears in the exponential due to the matrix multiplication of mean and covariance. This phase dependence is introduced independently of the time dynamics, as $f(t)\in\mathbb{R}$ ensures that no phase accumulation occurs in the primitive. The three functions defining the root-fidelity are derived as
\begin{align}
    \mathrm{F}(|r_{\beta}|,t) &= 2(16\cosh^2{|r_{\beta}|} -f^4(t)\sinh^2{|r_{\beta}|})^{-1/4} \\
    \mathrm{A}(|\beta|,|r_{\beta}|,t) &= (\frac{|\beta|}{8}(f(t)-2)[\mathrm{F(|r_{\beta}|,t)}]^2)^2 \\
    \mathrm{B}(|r_{\beta}|,\Theta,\Phi,t) &= 2(\cosh^2{|r_{\beta}|}[f^2(t)+4]\nonumber\\&\times(1+\cos[2\Theta-\Phi]\tanh{|r_{\beta}|})-f^2(t)).
\end{align}
We end this appendix by confirming that when $f(t) = 2$, we recover: $ \mathrm{F}(|r_{\beta}|,t) = 1$, $\mathrm{A}(|\beta|,|r_{\beta}|,t) = 0$, and $\mathrm{B}(|r_{\beta}|,\Theta,\Phi,t) =  16\cosh^2|r_\beta|(1+\cos[2\Theta-\Phi]\tanh|r_\beta|)-8$. This informs \eqref{eq:fidelity_struct} such that $\mathcal{F}(|\beta|,|r_{\beta}|,\Theta,\Phi,t) = 1\,e^{0\times\mathrm{B}(|r_{\beta}|,\Theta,\Phi,t)} = 1$ for all state parameters. Therefore, in the weak coupling regime, we find $f(t^*) \approx 2$ implies $\mathcal{F}(|\beta|,|r_{\beta}|,\Theta,\Phi,t) \approx 1$. The deviations of $f(t^*)$ from 2 are magnified in the Gaussian fidelity by the particular state parameters, as governed by $\mathrm{F}(|r_{\beta}|,t)$, $\mathrm{A}(|\beta|,|r_{\beta}|,t)$, and $\mathrm{B}(|r_{\beta}|,\Theta,\Phi,t)$. This is further amplified by the reported squared-fidelity, $\mathcal{F}^2$, as is seen in Sec.~\ref{subsec:Results_Baseline}.

\section{$\hat{O}_{0}$ Operator Definition and Derivation}
\label{app:oOperator}

This appendix provides the technical derivations for introducing finite memory effects via the O-zero methodology, as simulated in Sec.~\ref{subsec:Results_OU_Asymmetries}. We repeat that all other system and environment parameters are in the symmetric configuration, $\delta_{ab}=\eta=\epsilon=\delta\phi=0$, and the environment is now described by the single pole OU correlation in Eq. \ref{eq:OU_correlation_function}. The system dynamics are governed by the system Hamiltonian, bath coupling, and adjoint master equation defined in Eqs.~\eqref{eq:Hsys_bright_dark}, \eqref{eq:L_collective_dissipation}, and \eqref{eq:OU_ME}.

\subsection{Defining the $\hat{O}_0$ form and boundary conditions}

The O-zero method as defined by Yu \emph{et. al}\cite{yu_non-markovian_1999}, is centered around a time non-local operator. This operator is denoted by $\hat{O}_0(t,s)$, which encodes the non-local memory effects that can be defined by any combination of system operators that satisfy a predefined boundary and consistency condition. Part of the work reports a prior $\hat{O}_0(t,s)$ that satisfies both as 
\begin{equation}\label{eq:O0_definition}
    \hat{O}_0(t,s) = \sum_{o\in[m_\pm,n_{\pm}]}f_o(t,s)\hat{o} + g_o(t,s)\hat{o}^{\dagger}.
\end{equation}
The linear form is valid as the bilinear Hamiltonian, paired with the linear collective environmental coupling, allows the commutator algebra to be closed exactly for $o\in[m_\pm,n_{\pm}]$. The boundary conditions of this operator are defined by the appropriate recovery of the Lindblad coupling operator, $\hat{O}_0(t,t) = \hat{L}$. By aligning Eq.~\eqref{eq:O0_definition} and \eqref{eq:L_collective_dissipation}, which gives $\hat{L} = \sqrt{2\kappa}\hat{m}_{+}$ in the symmetry configuration, this is equivalent to 
\begin{align}\label{eq:O0_boundary_conditions}
    f_{m_{+}}(t,t) &= \sqrt{2\kappa} \\
    f_{o}(t,t) &= 0
    \\
    g_{o}(t,t) &= 0.
\end{align}
for all $o\in\{n_{\pm},m_{-}\}$ in the second and third lines.

\subsection{Consistency equation and partial derivatives}
\label{app:oOperator_consistency}

The second criterion requires the form of $\hat{O}_0(t,s)$ to satisfy the adjoint O-operator consistency equation for linear $\hat{L}$ and bilinear $\hat{H}_{\mathrm{sys}}'$, following Yu \emph{et. al}\cite{yu_non-markovian_1999} as
\begin{equation}\label{eq:O0_consistency_equation}
    \tfrac{\partial}{\partial t}\hat{O}_{0}(t,s) = -\tfrac{i}{\hbar}[\hat{H}_{\mathrm{sys}}',\hat{O}_0(t,s)] - [\hat{L}^{\dagger}\hat{\bar{O}}_0(t),\hat{O}_0(t,s)].
\end{equation}
This form is equivalent to the O-zero truncation of the time-nonlocal QSD equation under the replacement $\hat{O}(t,s)\to\hat{O}_0(t,s)$. This requirement introduces $\hat{\bar{O}}_0(t)$, which is $\hat{O}_0(t,s)$ dressed by the environmental memory function. Using the prior definition of $\hat{O}_0(t,s)$ we define 
\begin{align}\label{eq:O0_bar_definition}
    \hat{\bar{O}}_0(t) &= \int_{0}^{t}\alpha(t,s)\hat{O}_0(t,s)ds\nonumber \\
    &= \sum_{o\in[m_\pm,n_{\pm}]}F_o(t)\hat{o} + G_o(t)\hat{o}^{\dagger},
\end{align}
with dressed coefficients 
\begin{align}\label{eq:O0_bar_coefficients}
    F_o(t) &= \int_0^t\alpha(t,s)f_o(t,s)ds \\
    G_o(t) &= \int_0^t\alpha(t,s)g_o(t,s)ds.
\end{align}
The consistency equation then becomes a set of partial differential equations combining the coefficients of $\hat{\bar{O}}_0(t)$ and $\hat{O}_0(t,s)$. We recover
\begin{align}
    \frac{\partial f_o}{\partial t} &= \sqrt{2\kappa}\,F_o\,f_{m_{+}} +i\,g_a\,f_{u}\label{eq:O_f} \\
    \frac{\partial g_o}{\partial t} &= \sqrt{2\kappa}\,G_o\,f_{m_{+}} -i\,g_a\,g_{u}\label{eq:O_g}\\
    \frac{\partial g_{m_{+}}}{\partial t} &=  -i\,g_a\,g_{n_{+}} +\sqrt{2\kappa}\sum_{v\in\{n_{\pm},m_{-}\}}(f_vG_v-F_vg_v) \nonumber \\
    &+\sqrt{2\kappa}(2f_{m_+}G_{m_{+}}-F_{m_+}g_{m_{+}})\label{eq:O_gmp}
\end{align}
with $(o,u)\in \{(m_{\pm},n_{\pm}),(n_{\pm},m_{\pm})\}$ for the $f_o$ partial differential equations and $(o,u)\in \{(m_{-},n_{-}),(n_{\pm},m_{\pm})\}$ for $g_o$. At this point we emphasize Eqs.~\eqref{eq:O_f}-\eqref{eq:O_gmp} are not simulated directly as the dynamics only depends on $\hat{\bar{O}}_0(t)$. Instead, we solve the time-local ODEs for the coefficients $F_o$ and $G_o$ using these partial derivatives.

\subsection{Single-pole OU kernel and reduction to ODEs}
\label{app:oOperator_Obar}

For the Ornstein-Uhlenbeck correlation function in Eq.~\eqref{eq:OU_correlation_function}, the relationship of the coefficients in Eq.~\eqref{eq:O0_bar_coefficients} reduces to closed evolution equations,
\begin{align}
    \tfrac{d}{dt}F_o(t) = \frac{\gamma}{2}f_o(t,t&)-\tilde{\gamma}F_{o}(t)\nonumber\\&+\int_0^t\alpha_\mathrm{O}(t,s)\tfrac{\partial}{\partial t}f_o(t,s)ds\label{eq:Fo_generic_ODE}\\
    \tfrac{d}{dt}G_o(t) = \frac{\gamma}{2}g_o(t,t&)-\tilde{\gamma}G_{o}(t)\nonumber\\&+\int_0^t\alpha_\mathrm{O}(t,s)\tfrac{\partial}{\partial t}g_o(t,s)ds\label{eq:Go_generic_ODE}
\end{align}
with initial conditions $F_{o}(0)=G_{o}(0)=0$, and we define the complex-valued rate 
\begin{equation}\label{eq:OU_gamma_tilde}
    \tilde{\gamma} = \gamma + i\delta_{\mathrm{env}}.
\end{equation}
The last term is an implicit function of the desired coefficients ($F_o$ and $G_o$). Substituting Eqs.~\eqref{eq:O_f}-\eqref{eq:O_gmp} into Eqs.~\eqref{eq:Fo_generic_ODE}-\eqref{eq:Go_generic_ODE} and distributing the OU kernel produces integrals proportional to $F_o$ and $G_o$ only. We arrive at the following closed time-local system:
\begin{align}
    \tfrac{d}{dt}F_o &= -(\tilde{\gamma}-\sqrt{2\kappa}F_{m_+})F_o + ig_aF_u \label{eq:Fo_ODEs} \\
    \tfrac{d}{dt}G_o &= -(\tilde{\gamma}-\sqrt{2\kappa}F_{m_+})G_o - ig_aG_u \label{eq:Go_ODEs} \\
    \tfrac{d}{dt}F_{m_+} &=\frac{\gamma\sqrt{\kappa}}{\sqrt{2}} -(\tilde{\gamma}-\sqrt{2\kappa}F_{m_+})F_{m_+} + F_{n_+} \label{eq:Fmp_ODE},
\end{align}
where $(o,u)\in \{(m_{\pm},n_{\pm}),(n_{\pm},m_{\pm})\}$ for the $G_o$ partial differential equations,  $(o,u)\in \{(m_{-},n_{-}),(n_{\pm},m_{\pm})\}$ for $F_o$, and the initial conditions $F_o(0) = 0 = G_o(0)$ with $o\in[n_{\pm},m_{\pm}]$.

Inspecting this equation set, Eq.~\eqref{eq:Fmp_ODE} is the only non-homogeneous differential equation with a source term $\gamma\sqrt{\kappa/2}$ and will therefore act as the driver of the environmental dynamics. Moving to Eq.~\eqref{eq:Go_ODEs}, this corresponds to a linear homogeneous system for $G_o$ with $G_o(0) = 0$. By uniqueness of solutions for linear ODEs, $G_o(t) \equiv 0$ for all $t$ and no creation-operators appear in $\hat{\bar{O}}_0(t)$. Additionally, Eq.~\eqref{eq:Fo_ODEs} implies that $F_{m_-}$ and $F_{n_-}$ form a subset which forms the same linear homogeneous system as $G_o$. Given the same initial condition, uniqueness again forces $F_{m_-}(t) = F_{n_-}(t) = 0$ for all $t$. Lastly, $F_{m_+}$ acts as a time-dependent source term for $F_{n_+}$, leaving only the bright-sector coefficients to define the environmental effects. 

\subsection{Final coefficient ODEs}
\label{app:oOperator_finalODEs}

Applying the identities derived above, the OU environmental dynamics reduce to a set of two ODEs for the remaining coefficients,
\[
\{F_{n_{+}}(t),\,F_{m_{+}}(t)\},
\]
with explicit forms
\begin{align}
    \tfrac{d}{dt}F_{n_{+}} = &-\tilde{\gamma}F_{n_{+}}+\sqrt{2\kappa}F_{m_{+}}F_{n_{+}}+ig_aF_{m_{+}}\label{eq:ODE_Fnp}\\
    \tfrac{d}{dt}F_{m_{+}} = &\frac{\gamma\sqrt{\kappa}}{\sqrt{2}}-\tilde{\gamma}F_{m_{+}}+\sqrt{2\kappa}F_{m_{+}}^2+ig_aF_{n_+}\label{eq:ODE_Fmp},
\end{align}
and initial conditions $F_{m_+}(0) = F_{n_+}(0) = 0$. The O-zero dressed operator therefore reduces to the bright-sector subspace:
\begin{equation}\label{eq:O0_bar_bright_representation}
    \hat{\bar{O}}_0(t) = F_{m_+}(t)\hat{m}_{+} + F_{n_+}(t)\hat{n}_{+},
\end{equation}
as all remaining coefficients vanish identically. Eq.~\eqref{eq:ODE_Fnp}-\eqref{eq:ODE_Fmp} represent the reduced ODE set which is solved numerically, alongside Eq.~\eqref{eq:OU_ME}, to generate the results reported in Sec.~\ref{subsec:Results_OU_Asymmetries}. Additionally, as is standard for O-zero approximations, the resulting master equation is not in general of Lindblad form and complete positivity is no guaranteed; physicality is enforced numerically via the covariance bona fide condition as defined in Sec.~\ref{subsec:numerics}. Perturbative steady state solutions of these equations define the effective parameters found in the fast memory regime as derived in Appendix ~\ref{app:ouRates} and mentioned Sec.~\ref{subsec:open_system_models}.

\section{OU bright-sector renormalization in the fast memory regime}
\label{app:ouRates}

This appendix summarizes the perturbative steady-state procedure used to obtain the OU-renormalized bright-mode rates reported in Method~\ref{subsec:open_system_models}. The derivation starts from the reduced time-local coefficient equations of Appendix~\ref{app:oOperator}, where symmetry restricts the dynamics of the environment to the bright-sector defined by $F_{m_+}(t)$ and $F_{n_+}(t)$.

\subsection{Weak-coupling expansion}
\label{app:ouRates_expansion}

In the primitive operating regime $g_a/\kappa \ll 1$, we introduce the perturbative expansion
\begin{equation}
F_{o} = F_{o}^{(0)} + (\frac{g_a}{\kappa})F_{o}^{(1)} + (\frac{g_a}{\kappa})^2F_{o}^{(2)} + \mathcal{O}[(g_a/\kappa)^3],
\end{equation}
for $o\in\{n_{+},m_{+}\}$. Substitution into Eqs.~\eqref{eq:ODE_Fnp}-\eqref{eq:ODE_Fmp} yields a hierarchy of linear equations ordered by powers of $g_a/\kappa$, with 
\begin{itemize}
    \item \(F_{n_{+}}^{(0)}=0\), \(F_{n_{+}}^{(2)}=0\), and \(F_{m_{+}}^{(1)}=0\) due to the same homogeneous linear system argument as previous;
    \item the zeroth-order bright coefficient \(F_{m_{+}}^{(0)}(t)\) obeys a closed Riccati equation independent of higher-order terms and sets the effective bright-mode dynamics;
    \item \(F_{m_{+}}^{(0)}(t)\) acts as a time dependent source for \(F_{n_{+}}^{(1)}(t)\), which in turn acts as a time dependent source for \(F_{m_{+}}^{(2)}(t)\).
\end{itemize}
At each order, the equations take the form
\begin{align}
    \tfrac{d}{dt}F_{n_{+}}^{(1)} &= -\tilde{\gamma}F_{n_{+}}^{(1)} + \sqrt{2\kappa}F_{m_{+}}^{(0)}F_{n_{+}}^{(1)}+i\kappa F_{m_{+}}^{(0)}\label{eq:OU_Fnp1} \\
    \tfrac{d}{dt}F_{m_{+}}^{(0)} &= \frac{\gamma\sqrt{\kappa}}{\sqrt{2}}-\tilde{\gamma}F_{m_{+}}^{(0)} + \sqrt{2\kappa}(F_{m_{+}}^{(0)})^2\label{eq:OU_Fmp0}
    \\
    \tfrac{d}{dt}F_{m_{+}}^{(2)} &= -\tilde{\gamma}F_{m_{+}}^{(2)} + 2\sqrt{2\kappa}F_{m_{+}}^{(0)}F_{m_{+}}^{(2)} +i\kappa F_{n_+}^{(1)}.\label{eq:OU_Fmp2}
\end{align}
with initial conditions 
\begin{equation}\label{eq:OU_perturb_inits}
    F_{n_+}^{(1)}(0) = F_{m_{+}}^{(0)}(0) = F_{m_{+}}^{(2)}(0) = 0.
\end{equation}

\subsection{Zeroth-order Riccati solution and stability}
\label{app:ouRates_Riccati}

The zeroth-order equation for $F_{m_{+}}(t)$, Eq.~\ref{eq:OU_Fmp0}, admits an analytic Riccati-type solution,
\begin{equation} \label{eq:OU_Fmp0_solution}
    F_{m_{+}}^{(0)}(t) = (8\kappa)^{-1/2}(\tilde{\gamma}+\gamma_{\mathrm{O}}\tan[\tfrac{1}{2}\gamma_{\mathrm{O}}\,t+\Psi]),
\end{equation}
where 
\begin{align}
    \gamma_{\mathrm{O}} &= \sqrt{4\gamma\kappa-\tilde{\gamma}^2}\label{eq:OU_gammaO} \\
    \Psi &= \arctan[\tilde{\gamma}/\gamma_{\mathrm{O}}].\label{eq:OU_psi}
\end{align}
For $\delta_{\mathrm{env}}=0$ and $\gamma < 4\kappa$, this solution exhibits finite-time divergence, which is a feature of the single-pole OU kernel that signals a breakdown of the O-zero truncation. It is also consistent with earlier analyses of OU environments~\cite{diosi_non-markovian_1998}. 

For $\delta_{\mathrm{env}} \neq 0 \,\lor\,\gamma>4\kappa$, it is apparent that $\gamma_{\mathrm{O}}\in\mathbb{C}$. In this regime, the solution is analytically continued to a complex-valued form that approaches a steady state value. We report the full steady-state solutions as
\begin{align}
    \lim_{t\to\infty}F_{n_{+}}^{(1)}(t) &=i\sqrt{\frac{\kappa}{2}}\frac{(\tilde{\gamma}\pm i\gamma_{\mathrm{O}})}{(\tilde{\gamma}\mp i\gamma_{\mathrm{O}})} \label{eq:OU_Fnp1_infty}\\ 
    \lim_{t\to\infty}F_{m_{+}}^{(0)}(t) &= \frac{\tilde{\gamma}\pm i\gamma_{\mathrm{O}}}{2\sqrt{2\kappa}} \label{eq:OU_Fmp0_infty} \\
    \lim_{t\to\infty}F_{m_{+}}^{(2)}(t) &= i\sqrt{\frac{\kappa^3}{2\gamma_\mathrm{O}^2}}\frac{(\tilde{\gamma}\pm i\gamma_{\mathrm{O}})}{(\tilde{\gamma}\mp i\gamma_{\mathrm{O}})} \label{eq:OU_Fmp2_infty}
\end{align}
with corrections of order $\mathcal{O}\!\left[(g_a/\kappa)^{3}\right]$. The sign choice enforces continuity with the Markovian limit, i.e., when  $\gamma\to\infty$ one recovers $\lim_{t\to\infty}F_{m_{+}}^{(n)}(t) = \delta_{n,0}\sqrt{\kappa/2}$ and $\lim_{t\to\infty}F_{n_+}^{(1)}(t) = 0$. The branch that recovers the Markovian limit is equivalently defined by the sign of the environmental detuning. The Markovian recovery is ensured when the $+$ of the $\pm$ corresponds to $\delta_{\mathrm{env}}\leq0$ and the remaining case is assigned to $\delta_{\mathrm{env}}>0$.

\subsection{Steady-state coefficients}
\label{app:ouRates_steady}

With analytical solutions, the "fast memory" regime can be naturally defined whenever  $\delta_{\mathrm{env}} \neq 0 \,\lor\,\gamma>4\kappa$ as
\begin{equation}\label{eq:fast_memory_limit}
    \tau_{\mathrm{bath}} = \mathrm{Im}[\gamma_{\mathrm{O}}]^{-1} \ll \tau_{\mathrm{transfer}}=g_a^{-1}.
\end{equation}
This criterion is determined via a linear stability of the steady-state coefficients, Eqs.~\eqref{eq:OU_Fnp1_infty}-\eqref{eq:OU_Fmp2_infty} as defined by their corresponding dynamics in Eqs.~\eqref{eq:OU_Fnp1}-\eqref{eq:OU_Fmp2}. One finds that the requirement for the stability of the perturbative solutions is defined by $\mathrm{Im}[\gamma_{\mathrm{O}}] > 0$ for the $\delta_{\mathrm{env}} < 0$ branch and $\mathrm{Im}[\gamma_{\mathrm{O}}] < 0$ for the $\delta_{\mathrm{env}} > 0$ branch, both branches satisfy this condition for any $\gamma,\kappa > 0$. The corresponding rate of convergence is defined by eigenvalues of the linear stability analysis, giving $|\mathrm{Im}[\gamma_{\mathrm{O}}]|$ for the $m_+$ solutions and $|\mathrm{Im}[\gamma_{\mathrm{O}}]| + \gamma$ for the $n_+$ solution. We therefore define the slower of the two as the timescale of the bath, i.e., $\tau_{\mathrm{bath}} = |\mathrm{Im}[\gamma_{\mathrm{O}}]|^{-1}$. In the main study, we confirm that the default parameters yield $|\mathrm{Im}[\gamma_{\mathrm{O}}]|/g_a\approx 12.5 \gg 1$ which, combined with the $\delta_{\mathrm{env}}/\kappa = 0.6$, satisfy Eq.~\ref{eq:fast_memory_limit}. In summary, when both (i) a steady state exists ($\delta_{\mathrm{env}} \neq 0 \,\lor\,\gamma>4\kappa$) and (ii) the bath relaxation is much faster than the system dynamics ($\mathrm{Im}[\gamma_{\mathrm{O}}]^{-1} \ll g_a^{-1}$), the coefficients of $\hat{\bar{O}}_0(t)$ become effectively time independent on the transfer time scale as
\begin{align}
    F_{m_{+}}(\infty) &=\frac{\tilde{\gamma}\pm i\gamma_{\mathrm{O}}}{2\sqrt{2\kappa}}(1\mp\frac{2ig_a^2}{\gamma_{\mathrm{O}}(\tilde{\gamma}\mp i\gamma_{\mathrm{O}})}) + \mathcal{O}[(g_a/\kappa)^3]\label{eq:OU_Fmp_infty}\\ 
    F_{n_+}(\infty)  &= \frac{ig_a(\tilde{\gamma}\pm i\gamma_{\mathrm{O}})}{2\sqrt{\kappa}(\tilde{\gamma}\mp i\gamma_{\mathrm{O}})} + \mathcal{O}[(g_a/\kappa)^3]\label{eq:OU_Fnp_infty}.
\end{align}

\subsection{OU-renormalized rates}
\label{app:ouRates_effective}

In the combined regime of fast memory and weak-coupling, the form in Eq.~\eqref{eq:O0_bar_bright_representation} becomes approximately time-independent and solely updates the dynamics of the bright photon mode, with first-order moment
\begin{align}
    \tfrac{d}{dt}\langle \hat{m}_{+}\rangle = -&i[g_a-i\sqrt{2\kappa}F_{n_+}(\infty)]\langle\hat{n}_{+}\rangle\nonumber \\ &-[\sqrt{2\kappa}F_{m_+}(\infty)]\langle\hat{m}_{+}\rangle. \label{eq:OU_steady_bright_photon}
\end{align}
The terms are grouped to mirror the form of Eq.~\eqref{eq:first_order_ODEs}. Since both $F_{n_+}(\infty)$ and $F_{m_+}(\infty)$ are time independent, the bracketed terms represent renormalization of the coupling to the symmetric phonon mode and bright photon decay. Using the closed-form equations of the steady state coefficients, we recover 
\begin{align}\label{eq:OU_effective_rates_app}
    \kappa_{\mathrm{O}} &= \frac{\tilde{\gamma} \pm i\gamma_{\mathrm{O}}}{2}(1\mp\frac{2ig_a^2}{\gamma_{\mathrm{O}}(\tilde{\gamma} \mp i\gamma_{\mathrm{O}})})\\
    g_{\mathrm{O}} &= \frac{2g_a\tilde{\gamma}}{\tilde{\gamma} \mp i\gamma_{\mathrm{O}}}.
\end{align}
These parameters define the effective bright photon decay scale $\kappa_{\mathrm{eff}}=\mathrm{Re}[\kappa_{\mathrm{O}}]$, the effective bright coupling scale $g_{\mathrm{eff}}=\mathrm{Re}[g_{\mathrm{O}}]$, and the associated bright-channel rate 
\begin{equation}\label{eq:OU_gtilde_eff}
    \tilde{g}_{\mathrm{O}} = \sqrt{4g_ag_{\mathrm{O}}-\kappa_{\mathrm{O}}^2},
\end{equation}
which comes from redefining the second-order ODE seen in Eq.~\eqref{eq:first_order_bright_phonon_ODE}. The OU-modified interference function is therefore obtained by the substitutions $\kappa\rightarrow\kappa_{\mathrm{O}}$ and $\tilde g\rightarrow\tilde g_{\mathrm{O}}$ in the Markovian bright-channel solution $\mathrm{T}_{+}(t)$ of Eq.~\eqref{eq:first_order_bright_phonon}. We note that this procedure and substitution are strictly valid in the previously defined "fast memory" regime. All reported OU results use the full O-zero time-local ODEs given by Eqs.~\eqref{eq:ODE_Fnp} and \eqref{eq:ODE_Fmp} with results shown in Sec.~\ref{subsec:Results_OU_Asymmetries}. The steady-state expressions are used only for interpretation within the fast-memory regime.

\section{Asymmetry Derivations}\label{app:Markov_Asymmetries}

This appendix derives the analytic Markovian dynamics for the two asymmetries that preserve closed-form bright–dark block structure: the relative laser phase $\delta\phi$ and the phonon–photon detuning $\delta_{ab}$. All other asymmetries considered in the main text—coupling imbalance $\eta$, and decay-rate asymmetry $\epsilon$—induce bright–dark mixing and therefore require numerical treatment.

Throughout, we work with the dynamics defined by the baseline primitive as defined in Sec.~\ref{subsec:collective_dissipation} and derived in Appendix~\ref{app:Markov_Exact}.

\subsection{Relative coupling phase \texorpdfstring{$\delta\phi$}{δφ}}
\label{app:coupling_phase}

Introducing a relative phase between the interaction terms, the native phonon basis $(\hat b,\hat d)$ absorbs this contribution and breaks the original symmetric-antisymmetric collective phonon definition. This updates Eq.~\eqref{eq:Hsys_bright_dark} to a form where only the photon modes are decomposed into their collective counterparts as 
\begin{equation}\label{eq:Hsys_couplingPhase}
     \hat{H}_{\mathrm{sys}}'/\hbar =  g_a(\hat{m}_{-}[\hat{b}^{\dagger}-e^{-i\,\delta\phi}\hat{d}^{\dagger}]+ \hat{m}_{+}[\hat{b}^{\dagger}+e^{-i\,\delta\phi}\hat{d}^{\dagger}]) +\mathrm{h.c.}
\end{equation}
where the factor $e^{-i\,\delta\phi}$ multiplies only $\hat d$. The block structure is restored by the phase-rotated symmetric-antisymmetric collective operators
\begin{equation}\label{eq:collective_phonon_phase}
    \hat{n}_{\pm,\delta\phi} = \frac{1}{\sqrt{2}}(\hat{b} \pm e^{i\delta\phi}\hat{d}).
\end{equation}
This transformation is unitary and preserves the bosonic commutation relations, ensuring that the resulting bright–dark phonon modes form an orthonormal collective basis. Additionally, the Hamiltonian recovers the form of Eq.~\eqref{eq:Hsys_bright_dark} with $\hat{n}_{\pm} \to \hat{n}_{\pm,\delta\phi}$ and the bright and dark sectors evolve identically to the symmetric case. The adjoint master equation~\eqref{eq:Markov_ME} therefore yields the same dynamic systems as Eqs.~\eqref{eq:first_order_ODEs}, and the channel amplitudes $\mathrm{T}_{\pm}(t)$ remain unchanged.

The modification enters only through transfer of the physical target mode. Inverting the phase-rotated basis gives 
\begin{equation}\label{eqd_transfer_phase_case}
    \hat{d} = \frac{1}{\sqrt{2}}e^{-i\delta\phi}(\hat{n}_{+,\delta\phi}-\hat{n}_{-,\delta\phi})
\end{equation}
which introduces an overall phase factor in all first- and second-order target moments without affecting their magnitudes or relative bright–dark weights. The resulting time-dependence takes the form
\begin{equation}\label{eq:f_phase_case}
    f_{\delta\phi}(t) = e^{-i\delta\phi}[e^{-\kappa\,t/2}(\cosh[\tfrac{1}{2}\tilde{g}\,t] + \tfrac{\kappa}{\tilde{g}}\sinh[\tfrac{1}{2}\tilde{g}\,t]) - \cos[g_a\,t]],
\end{equation}
with the bracketed factor equal to the symmetric time-dependence function of Eq.~\eqref{eq:exact_target_first_order_moment}. Because the relative phase enters only through a unitary rotation of the collective phonon basis, it introduces no dynamical modification to the bright or dark blocks and therefore alters only the transfer map to the physical target mode. As reported in Sec.~\ref{subsec:Results_Markov_Asymmetries}, the primitive becomes highly sensitive to this asymmetry for even a small relative phase. This is explained by the strong influence this asymmetry has on the state misalignment $\mathrm{B}(|r_\beta|,\Theta,\Phi,t)$ penalty in the fidelity metric. We emphasize that this effect is purely a \emph{deterministic} phase rotation, as $|f_{\delta\phi}(t)| = f(t)$ and the phase is determined by experimentally available parameters unrelated to the state information. This implies that experimental controls can be used to undo the effects of this asymmetry, which is beyond the scope of this study.

\subsection{Phonon–photon detuning \texorpdfstring{$\delta_{ab}$}{δab}}
\label{app:MarkovAsym_detuning}

Introducing a finite phonon–photon detuning $\delta_{ab}$ adds self-energy terms to the Hamiltonian in Eq.~\eqref{eq:Hsys_bright_dark} for both collective phonon modes in the bright–dark basis, yielding
\begin{equation}\label{eq:Hsys_bright_dark_detuning}
    \hat{H}' = \hat{H}_{\mathrm{sys}}' + \hbar\delta_{ab}(\hat{n}_{+}^{\dagger}\hat{n}_{+} + \hat{n}_{-}^{\dagger}\hat{n}_{-}). 
\end{equation}
This preserves the block structure of the symmetric case while shifting the local energies of $\hat n_{+}$ and $\hat n_{-}$.

Substituting this Hamiltonian into the Markovian adjoint equation leaves the bright-dark photon dynamics unchanged (i.e., governed by Eq.~\eqref{eq:first_order_ODEs}) and updates the phonon dynamics as
\begin{align}\label{eq:photon_ODEs_detuning}
    \tfrac{d}{dt}\langle \hat{n}_{-}\rangle &= -ig_a\langle \hat{m}_{-}\rangle - i\delta_{ab}\langle \hat{n}_{-}\rangle \\ \tfrac{d}{dt}\langle \hat{n}_{+}\rangle &= -ig_a\langle \hat{m}_{+}\rangle - i\delta_{ab}\langle \hat{n}_{+}\rangle.
\end{align}
with initial conditions identical to Appendix~\ref{app:Markov_Exact}. The dark sector remains closed and oscillates at the detuning-shifted frequency
\begin{equation}\label{eq:g_delta_detuning}
    g_{\delta} = \sqrt{4g_a^2+\delta_{ab}^2},
\end{equation}
which is the eigenfrequency of the closed dark-sector block, i.e., the normal-mode splitting of the two-mode real Hamiltonian generated by excitation converting interaction in the presence of detuning. In addition, the dark sector acquires an overall phase that is applied to the quantum information transferred via this channel. The corresponding dark-mode amplitude is
\begin{equation}\label{eq:dark_detuning}
    \langle \hat{n}_{-}\rangle = \frac{\beta}{\sqrt{2}}e^{-i\delta_{ab}\,t/2}(\cos[\tfrac{1}{2}g_{\delta}\,t] - \frac{i\delta_{ab}}{g_{\delta}}\sin[\tfrac{1}{2}g_{\delta}\,t]).
\end{equation}
The bright sector acquires both detuning and damping, characterized by the complex rate
\begin{equation}\label{eq:g_tilde_detuning}
    \tilde{g}_{\delta} = \sqrt{(\kappa-i\,\delta_{ab})^2-4g_a^2}
\end{equation}
which arises from the characteristic equation of the bright sector. The solution is provided as
\begin{equation}\label{eq:bright_detuning}
    \langle \hat{n}_{+}\rangle = \frac{\beta}{\sqrt{2}}e^{\frac{-(\kappa+i\delta_{ab})t}{2}}(\cosh[\tfrac{1}{2}\tilde{g}_{\delta}\,t] + \frac{\kappa-i\delta_{ab}}{\tilde{g}_{\delta}}\sinh[\tfrac{1}{2}\tilde{g}_{\delta}\,t]).
\end{equation}
Reconstructing the target-mode first-order moment via $\hat d=(\hat n_{+}-\hat n_{-})/\sqrt{2}$ yields
\begin{equation}\label{eq:d_mean_detuning}
    \langle \hat{d} \rangle_t = \frac{\beta}{2}f_{\delta_{ab}}(t)
\end{equation}
with the detuning-dependent interference function
\begin{multline}\label{eq:f_detuning_case}
   f_{\delta_{ab}}(t) = e^{\frac{-i\delta_{ab}\,t}{2}}[e^{\frac{-\kappa\,t}{2}}(\cosh[\tfrac{1}{2}\tilde{g}_{\delta}\,t] + \frac{\kappa-i\delta_{ab}}{\tilde{g}_{\delta}}\sinh[\tfrac{1}{2}\tilde{g}_{\delta}\,t])\\-(\cos[\tfrac{1}{2}g_{\delta}\,t] - \frac{i\delta_{ab}}{g_{\delta}}\sin[\tfrac{1}{2}g_{\delta}\,t])].
\end{multline}
which reduces to Eq.~\eqref{eq:exact_target_first_order_moment} for $\delta_{ab}=0$. For finite detuning, $f_{\delta_{ab}}(t)$ is generally complex and quickly deviates from the time evolution of the pure symmetric case. Unlike the previously mentioned phase offset, this asymmetry directly alters the effective dynamical rates and therefore the predicted transfer time ($t^*$). This prevents the deterministic state transfer defined by our protocol beyond extremely small detuning as reported in Sec.~\ref{subsec:Results_Markov_Asymmetries}.

\subsection{Other asymmetries: numerical treatment}
\label{app:MarkovAsym_other}

The remaining system asymmetries ($\eta$,$\epsilon$) do not preserve the analytic bright–dark decomposition. The coupling asymmetry ($\eta$) breaks the bright-dark symmetry on the Hamiltonian level, whereas the decay asymmetry ($\epsilon$) does so at the environment level. 

Each breaks the block-diagonal structure underlying Eqs.~\eqref{eq:first_order_ODEs}, precluding closed-form expressions for the interference function $f(t)$. Their effects are therefore obtained from direct numerical integration of the full first- and second-order moment equations following Method~\ref{subsec:numerics}. The quantitative roles of each imperfection and their impacts on the dissipative transfer primitive are summarized in Result~\ref{subsec:Results_Markov_Asymmetries}.

\bibliography{DSR_Fixed_Time_Gaussian_State_Transfer_via_Collective_Dissipation}

\end{document}